\documentclass[traditabstract]{aa} 
\usepackage{graphicx}
\usepackage{txfonts}
\usepackage{lscape}
\usepackage{longtable}
\usepackage{natbib}

\newcommand{\micron}{$\mu$m}

\newcommand{\oia}{[\ion{O}{i}]\,63\,\micron}
\newcommand{\oib}{[\ion{O}{i}]\,145\,\micron}
\newcommand{\cii}{[\ion{C}{ii}]}
\newcommand{\texi}{$T_{\rm ex}$}
\newcommand{\ncol}{$N_{\rm mol}$}
\newcommand{\column}{cm$^{-2}$}
\newcommand{\density}{cm$^{-3}$}

\newcommand{\redchis}{${\tilde\chi}^2$}
\newcommand{\herschel}{{\it Herschel}}

\begin{document}

   \title{DIGIT survey of far-infrared lines from protoplanetary disks I. [\ion{O}{i}], [\ion{C}{ii}], OH, H$_2$O and CH$^+$.
   \thanks{{\it Herschel} is an ESA space observatory with science instruments
provided by European-led Principal Investigator consortia and with important
participation from NASA.}}

   \author{
	D.\ Fedele\inst{\ref{inst_mpe}},
        S.\ Bruderer\inst{\ref{inst_mpe}},
        E.F.\ van Dishoeck\inst{\ref{inst_mpe},\ref{inst_leiden}},
        J. \ Carr\inst{\ref{inst_navy}},
        G.J.\ Herczeg\inst{\ref{inst_kavli}},
        C.\ Salyk\inst{\ref{inst_noao}}
        Neal J.\ Evans II\inst{\ref{inst_austin}}, 
        J.\ Bouwman\inst{\ref{inst_mpia}}, 
        G.\ Meeus\inst{\ref{inst_madrid}},
        Th.\ Henning\inst{\ref{inst_mpia}}, 
        J.\ Green\inst{\ref{inst_austin}},
        J.R.\ Najita\inst{\ref{inst_noao}}
        M. \ Guedel\inst{\ref{inst_wien}}
}

\institute{
Max Planck Institut f\"{u}r Extraterrestrische Physik, Giessenbachstrasse 1, 85748 Garching, Germany\label{inst_mpe}
\and
Leiden Observatory, Leiden University, P.O. Box 9513, 2300 RA Leiden, The Netherlands\label{inst_leiden}
\and 
Naval Research Laboratory, Code 7211, Washington, DC 20375, USA\label{inst_navy}
\and
Kavli Institute for Astronomy and Astrophysics, Yi He Yuan Lu 5, Beijing, 100871, P.R. China\label{inst_kavli}
\and
National Optical Astronomy Observatory, 950 N. Cherry Avenue, Tucson, AZ 85719, USA\label{inst_noao}
\and
University of Texas at Austin, Department of Astronomy, 2515 Speedway,
Stop C1400, Austin TX 78712-1205, USA\label{inst_austin}
\and
Max Planck Institute for Astronomy, K\"onigstuhl 17, 69117, Heidelberg, Germany\label{inst_mpia}
\and
Universidad Autonoma de Madrid, Dpt. Fisica Teorica, Campus Cantoblanco, Spain\label{inst_madrid}
\and
Universit\"at Wien, Dr.-Karl-Lueger-Ring 1, 1010, Wien, Austria\label{inst_wien}
}


\authorrunning{Fedele et al.}
\offprints{Davide Fedele,\\ \email{fedele@mpe.mpg.de}}
 
  \abstract{
    We present far-infrared (50-200\,\micron) spectroscopic
    observations of young pre-main-sequence (PMS) stars taken with \herschel/PACS 
    as part of the DIGIT
    key project. The sample includes 16 Herbig AeBe and 4 T Tauri
    sources observed in SED mode covering the entire spectral
    range. An additional 6 Herbig AeBe and 4 T Tauri systems have been
    observed in SED mode with a limited spectral coverage. Multiple
    atomic fine structure and molecular lines are detected at the source position:
    [\ion{O}{i}], [\ion{C}{ii}], CO, OH, H$_2$O, CH$^+$. The most common feature
    is the \oia \ line detected in almost all of the sources followed
    by OH. In contrast with CO, OH is detected toward both Herbig AeBe
    groups (flared and non-flared sources). 
    An isothermal LTE slab model fit to the OH lines indicates column densities  of 
    $10^{13} < N_{\rm  OH} < 10^{16}\,$\column, emitting radii $15 < r < 
    100$\,AU and excitation temperatures  $100 < T_{\rm ex} < 400\,$K. 
    We used the non-LTE code RADEX to verify the LTE assumption. High gas densities 
      ($n \geq 10^{10}\,$cm$^{-3}$) are needed to reproduce the observations.
    The OH emission thus comes from a warm layer in the disk at intermediate stellar 
    distances. Warm H$_2$O emission is detected through multiple lines toward the 
    T Tauri systems AS 205, DG Tau, S CrA and RNO 90 and three Herbig AeBe
    systems HD 104237, HD 142527, HD 163296 (through line stacking).
    Overall, Herbig AeBe sources have higher OH/H$_2$O
    abundance ratios across the disk than do T Tauri disks, from near- to far-infrared wavelengths.
    Far-infrared CH$^+$ emission is detected toward HD 100546 and HD 97048. The slab model 
    suggests moderate excitation (\texi \ $\sim$ 100\,K) and compact ($r \sim 60\,$AU) emission in 
    the case of HD 100546. Off-source [\ion{O}{i}] emission is detected toward DG Tau, whose origin is likely the 
    outflow associated with this source. The \cii \ emission is spatially extended in all sources where 
    the line is detected. 
    This suggests that not all \cii \ emission is associated with the disk and that there is a substantial
    contribution from diffuse material around the young stars. The flux ratios of the atomic fine structure 
    lines (\oia, \oib, \cii) are analyzed with PDR models and require high gas density ($n \sim 10^5$\,cm$^{-3}$) 
    and high UV fluxes ($G_o \sim 10^3 - 10^7$), consistent with a disk origin for the oxygen lines for  most of the 
    sources.  
    }
   \maketitle

\section{Introduction}

Far-infrared (far-IR) spectroscopic observations of young pre-main-sequence
stars have the potential to reveal the gas and dust composition of
protoplanetary disks in regions not probed at any other wavelengths
\citep[e.g.,][]{vanDishoeck04,Lorenzetti05,Henning10}. The atomic and
molecular transitions in the far-IR regime (50-200\,\micron)
span a large range in upper energy level (from a few 10\,K to a
few 10$^3$\,K) and are sensitive to the warm (a few 10$^2$\,K) upper
layers of the disk ($n < 10^8$\,\density). For a disk irradiated 
by UV and/or X-rays from the pre-main-sequence star, these conditions
are found at intermediate distances from the central star ($r \gtrsim$
10\,AU) \citep[e.g.,][]{Kamp04,Bruderer12}. Observations of lines of
multiple species provide a wealth of information that allow us to (1)
determine the physical properties of the gas such as excitation
temperature, column density, emitting radii (and in some cases the
total gas density); (2) constrain the excitation mechanism (e.g.,
collisions, UV fluorescence, IR pumping); and (3) address the chemical
structure of the disk. The far-IR (IR) spectrum contains
information complementary to that provided by near- and mid-IR
observations which are sensitive to the hot ($>$ 1000\,K) inner region
of the disk ($<$few AU). At the other end of the spectrum,
(sub)millimeter spectroscopic observations with ALMA will unveil the
physical conditions and chemical composition of the disk midplane at
distances $r \gtrsim$ 10\,AU. The far-IR data probe intermediate disk 
radii and depths. The ultimate goal of these
observational campaigns is to use the combined data to address the
chemistry and physics of the entire protoplanetary disk from inner to
outer edge.

\smallskip
\noindent
We present here 50--200\,\micron \ spectra of a sample of protoplanetary
disks around Herbig AeBe and T Tauri stars obtained in the context of
the `Dust, Ice and Gas in Time' (DIGIT) key program \citep{Sturm10}.  
The unprecedented sensitivity of the PACS instrument
\citep{Poglitsch10} on board the {\it Herschel} Space Observatory
\citep{Pilbratt10} allows for the first time the detection of weak atomic and
molecular emission down to a few 10$^{-18}$\,W\,m$^{-2}$. Far-IR
spectra of bright Herbig stars have been obtained previously with the
Long Wavelength Spectrometer (LWS) on the Infrared Space Observatory
(ISO) \citep[e.g.,][]{Waelkens96,Meeus01,Giannini99,Lorenzetti99,Lorenzetti02,Creech-Eakman02}. One
of the main results has been an empirical classification of the Herbig
AeBe systems into two groups based on the ratio of the far- to near-IR
(dust) emission \citep{Meeus01}. Group I sources have a high far-
to near-IR emission ratio consistent with a flaring disk geometry
while Group II sources have a low flux ratio characteristic of a
flat, self-shadowed disk.  Grain growth and settling may also play a
role \citep[e.g.][]{Acke09}.  One question to be addressed here is to
what extent the far-IR gas-phase lines reflect this dichotomy in disk
structure.

\smallskip
\noindent
The near-IR spectra of Herbig AeBe disks are characterised by
several ro-vibrational lines of CO \citep[e.g.][]{Brittain03,Blake04,vanderplas09, Salyk11b} 
and OH \citep{Mandell08, Fedele11,Doppmann11,Liskowsky12}. At mid-IR wavelengths the 
spectra of Herbig AeBe disks are dominated by dust emission and only  very few Herbig 
sources show molecular emission \citep{Pontoppidan10, Salyk11a}.  
The optical forbidden oxygen lines are common in Herbig AeBe spectra \citep[e.g.][]{Acke04} and
are found to come from the disk atmosphere close to the star \citep[$<10\,$AU, e.g.,][]{Fedele08, VanderPlas08}.
In contrast, the emission
from T Tauri systems is characterised by a rich molecular spectrum
from near- to mid-IR wavelengths.  The inventory of molecular species
detected in T Tauri sources in the infrared includes: CO
\citep[e.g.][]{Najita03}, OH and H$_2$O \citep[e.g.,][]{Carr04, Salyk08}, HCN and
C$_2$H$_2$ \citep[e.g.,][]{Carr08, Pascucci09, Carr11, Mandell12} and, finally CO$_2$
\citep{Pontoppidan10}. Are Herbig sources also different from T Tauri
sources at far-IR wavelengths?

\smallskip
\noindent
In this paper we report on the detection of far-IR atomic fine
structure lines ([\ion{O}{i}] and [\ion{C}{ii}]) and molecular lines
(OH, H$_2$O, CH$^+$). The analysis of far-IR CO lines is
reported in \citet[][hereafter paper II]{Meeus13}. 
This survey over the full PACS wavelength range complements GASPS \citep{Meeus12}
which targeted specific lines.

\begin{table*}[!ht]
\caption{Properties of the program stars and PACS observation log}
\label{tab:sample}   
\centering         
\begin{tabular}{lllllllllll}
\hline\hline            
Star        & RA     & DEC     & Sp.  Type & Distance\tablefootmark      &  Group & Obsid   & Obs. date \\
            & (J2000)& (J2000) &           & [pc]                        &        & (1342+) &     \\
\hline                                                       
AB Aur      & 04 55 45.8 & +30 33 04.3 & A0    & 140  $\pm$ 15     \tablefootmark{a}    & I  &217842/3                       & 2011/04/04 \\
HD 35187    & 05 24 01.2 & +24 57 37.6 & A2+A7 & 114  $\pm$ 24     \tablefootmark{a}    & II &217846\tablefootmark{\dagger}  & 2011/04/04 \\
HD 36112    & 05 30 27.5 & +25 19 57.0 & A5    & 280  $\pm$ 55     \tablefootmark{a}    & I  &228247/8                       & 2011/09/07 \\
HD 38120    & 05 43 11.9 & -04 59 49.9 & B9    & 480  $\pm$ 175    \tablefootmark{a}    & I  &226212/3                       & 2011/08/15 \\
HD 50138    & 06 51 33.4 & -06 57 59.5 & B9    & 390  $\pm$ 70     \tablefootmark{a}    & II &206991/2                       & 2010/10/23 \\
HD 97048    & 11 08 03.3 & -77 39 17.4 & A0    & 160  $\pm$ 15     \tablefootmark{a}    & I  &199412/3                       & 2010/06/30 \\
HD 98922    & 11 22 31.7 & -53 22 11.5 & B9    & 1150$^{+935}_{-355}$ \tablefootmark{a}   & II & 210385\tablefootmark{\ddagger} & 2010/11/27 \\
HD 100453   & 11 33 05.6 & -54 19 28.5 & A9    & 122  $\pm$ 10     \tablefootmark{a}    & I  &211695/6                       & 2010/12/25 \\
HD 100546   & 11 33 25.4 & -70 11 41.2 & B9    & 97   $\pm$ 4      \tablefootmark{a}    & I  &188038/7                       & 2009/12/11 \\
HD 104237   & 12 00 05.1 & -78 11 34.6 & A4    & 116  $\pm$ 5      \tablefootmark{a}    & II &207819/20                      & 2010/11/03 \\
HD 135344 B & 15 15 48.4 & -37 09 16.0 & F4    & 140  $\pm$ 27     \tablefootmark{b}    & I  &213921/2                       & 2011/02/07 \\
HD 139614   & 15 40 46.4 & -42 29 53.5 & A7    & 140  $\pm$ 5      \tablefootmark{c}    & I  &215683/4                       & 2011/03/10 \\
HD 141569 A & 15 49 57.8 & -03 55 16.3 & A0    & 116  $\pm$ 7      \tablefootmark{a}    & II &213913\tablefootmark{\ddagger} & 2011/02/06 \\
HD 142527   & 15 56 41.9 & -42 19 23.2 & F6    & 230  $\pm$ 50     \tablefootmark{a}    & I  &216174/5                       & 2011/03/16 \\
HD 142666   & 15 56 40.0 & -22 01 40.0 & A8    & 145  $\pm$ 5      \tablefootmark{c}    & II &213916\tablefootmark{\ddagger} & 2011/02/06 \\
HD 144432   & 16 06 57.9 & -27 43 09.7 & A9    & 160  $\pm$ 25     \tablefootmark{a}    & II &213919\tablefootmark{\ddagger} & 2011/02/07 \\
HD 144668   & 16 08 34.3 & -39 06 18.3 & A1/A2 & 160  $\pm$ 15     \tablefootmark{a}    & II &215641/2                       & 2011/03/08 \\
Oph IRS 48  & 16 27 37.2 & -24 30 35.0 & A0    & 120  $\pm$ 4      \tablefootmark{d}    & I  &227069/70                      & 2011/08/22 \\
HD 150193   & 16 40 17.9 & -23 53 45.2 & A2    & 203  $\pm$ 40     \tablefootmark{a}    & II &227068\tablefootmark{\ddagger} & 2011/08/22 \\
HD 163296   & 17 56 21.3 & -21 57 21.9 & A1    & 120  $\pm$ 10     \tablefootmark{a}    & II &217819/20                      & 2011/04/03 \\
HD 169142   & 18 24 29.8 & -29 46 49.3 & A8    & 145  $\pm$ 5      \tablefootmark{c}    & I  &206987/8                       & 2010/10/23 \\
HD 179218   & 19 11 11.3 & +15 47 15.6 & A0    & 255  $\pm$ 40     \tablefootmark{a}    & I  &208884/5                       & 2010/11/12 \\
\hline                                                                                                                                  
DG Tau      & 04 27 04.7 & +26 06 16.3 & K5    & 140               \tablefootmark{e}    &    &225730/1                       & 2011/11/15 \\
HT Lup      & 15 45 12.9 & -34 17 30.6 & K2    & 120  $\pm$ 35     \tablefootmark{a}    &    &213920\tablefootmark{\ddagger} & 2011/11/17 \\
RU Lup      & 15 56 42.3 & -37 49 15.5 & G5    & 120  $\pm$ 35     \tablefootmark{a}    &    &215682\tablefootmark{\ddagger} & 2011/03/10 \\
RY Lup      & 15 59 28.4 & -40 21 51.2 & K4    & 120  $\pm$ 35     \tablefootmark{a}    &    &216171\tablefootmark{\ddagger} & 2011/03/16 \\
AS 205      & 16 11 31.4 & -18 38 24.5 & K5    & 125               \tablefootmark{f}    &    &215737/8                       & 2011/11/18 \\
EM* SR 21   & 16 27 10.3 & -24 19 12.5 & G3    & 120  $\pm$  4     \tablefootmark{f}    &    &227209/10                      & 2011/08/14 \\
RNO 90      & 16 34 09.2 & -15 48 16.8 & G5    & 125  $\pm$  4     \tablefootmark{f}    &    &228206\tablefootmark{\ddagger} & 2011/09/06 \\
S Cra       & 19 01 08.6 & -36 57 19.9 & K3+M0 & 129  $\pm$ 11     \tablefootmark{g}    &    &207809/10                      & 2010/11/02 \\
\hline                                   
\end{tabular}
\tablefoot{
\tablefoottext{a}{\citet{vanLeeuwen07}};
\tablefoottext{b}{\citet{Muller11}};
\tablefoottext{c}{\citet{Acke04}} and references therein;
\tablefoottext{d}{\citet{Loinard08}};
\tablefoottext{d}{\citet{Kenyon08}};
\tablefoottext{f}{\citet{Pontoppidan11}} and references therein;
\tablefoottext{g}{\citet{Neuhauser08}}\\
\tablefoottext{\dagger} {Spectral coverage = 50-73\,\micron, 100-145\,\micron;}
\tablefoottext{\ddagger} {Spectral coverage = 60-75\,\micron, 120-143\,\micron}
}
\end{table*}

\section{Observations and data reduction}
\subsection{Sample}
The sources were selected primarily on their far-IR fluxes such that a
$S/N\approx 100$ could be reached on the continuum within 5 hours of
integration time. 
The Herbig AeBe sources in this sample have spectral type between F4 to B9 and are not embedded in large 
molecular clouds. They have been studied previously at mid-IR wavelengths by {\it Spitzer} 
\citep{Juhasz10} and the selected sample contains mostly nearby and low-luminosity sources. 
The T Tauri stars consist of an inhomogeneous sample of bright sources with K--G 
spectral type. AS 205, S CrA, and RU Lup are heavily veiled sources, with CO line profiles 
suggesting the presence of a disc wind \citep{Bast11, Pontoppidan11}. DG Tau is associated 
with an outflow that can contribute to the observed emission. In addition RU Lup has evidence 
for a jet \citep{Guedel10}. Table~\ref{tab:sample} provides the parameters of the sample. 
For the Herbig AeBe sources, the disc group is also indicated: group I sources have flared discs while
group II sources have flat discs, in the classification of \citealt{Meeus01}).

\noindent
The focus in this paper is on the Herbig sample, but the data on T Tauri
sources are reported for completeness and to allow a comparison with
the Herbig sample in a consistent way. More details about the sample are 
given in paper II.

\subsection{Observational details}\label{sec:obs_1}
PACS is an array of 5$\times$5 
spaxels\footnote{A spaxel is a spatial sampling element of the PACS integral field unit}, 
with each spaxel covering
9\farcs4$\times$9\farcs4. The instrument is diffraction limited only
at $\lambda\,<\,110$\,\micron.  The targets were observed in SED mode
with two settings in order to cover the spectral range 51-220\,\micron
\ (B2A, 51-73\,$\mu$m, short R1, 100-145\,\micron \ and B2B,
70-105\,\micron \, long R1, 140-220\,\micron). The spectral resolving
power is $R=\lambda/\Delta \lambda \sim$ 1000, increasing to 3000 at
the shortest wavelengths. A second sample of targets was observed with
a limited spectral range (B2A, 60-75 \micron; short R1, 120-143 \micron)
centered at the position of the forsterite emission but including some
specific lines. The observations were carried out in chopping/nodding
mode with a chopping throw of 6\arcmin.  The observation log and
parameters of the sample are presented in Table~\ref{tab:sample}.

\smallskip
\noindent
The data have been reduced with HIPE 8.0.2489 with standard
calibration files from level 0 to level 2 \citep[see][]{Green13}. 
The two nod positions were reduced separately
(oversampling factor = 3) and averaged after a
flat-field correction. In the case of HD 100546, which was observed in
a different mode during the science demonstration phase, we used an
oversampling factor equal to 1. The spectra are extracted from the
central spaxel to optimize the signal-to-noise ($S/N$) ratio. To flux
calibrate the spectra we performed the following steps: 1) correct for
flux loss by means of a PSF-loss correction function provided by HIPE;
2) scale to PACS photometry (whenever available); 3) matching spectral
modules. Step 1 is valid for objects well centered in the central
spaxel. In the case of mispointed observations we extracted the total
flux (all 25 spaxels) to recover the flux loss. In this case we fitted
a 3$^{rd}$-order polynomial to two spectra (central spaxel and 25
spaxels). The correction factor is the ratio between the two fits. The
mispointed sources are: AB Aur, HD 97048, HD 169142, HD 142666. The
regions affected by spectral leakage (B2B $95 - 105\,\mu$m and R1
$190-220\,\mu$m) are excluded from this procedure. Based on a
statistical analysis, the PACS SED fluxes agree with PACS photometry
to within 5--10\%. For this reason we assign an uncertainty of 10\% to
the PACS SED fluxes of sources without PACS photometry available.

\smallskip
\noindent
The line fluxes are measured by fitting a Gaussian function and the 
uncertainty ($\sigma$) is given by the product $STD_F \ \delta \lambda \ \sqrt{N_{\rm bin}}$, 
where $STD_F$ is the standard deviation of the (local) spectrum (W\,m$^{-2}$\,\micron$^{-1}$), 
$\delta \lambda$ is the wavelength spacing of the bins (\micron) and $N_{\rm bin}$ is the width of the
line in spectral bins (5 for all lines).

\begin{table}
\caption{Overview of detected species. }
\label{tab:overview}   
\centering   
\resizebox{\hsize}{!} {          
\begin{tabular}{llllllll}
\hline\hline            
 Star                 & \multicolumn{2}{c}{$[\ion{O}{i}]$} & $[\ion{C}{ii}]$ & CO\tablefootmark{a} & OH & H$_2$O & CH$^+$  \\  
                      &  63\,$\mu$m                & 145\,$\mu$m                & & & & & \\
\hline
 AB Aur               & Y                         & Y                          & Y               & Y  & Y  & n      & n          \\	     	
 HD 35187             & Y                         & n\tablefootmark{b}         & n\tablefootmark{b}  & n  & n  & n      & n      \\
 HD 36112             & Y                         & n                          & n               & Y  & Y  & n      & n          \\	     	
 HD 38120             & Y                         & Y                          & Y               & n  & ?  & n      & n          \\	     	
 HD 50138             & Y                         & Y                          & Y               & n  & Y  & n      & n          \\	     	
 HD 97048             & Y                         & Y                          & Y               & Y  & Y  & n      & Y          \\	     	
 HD 98922             & Y                         & -                          & -               & n  & n  & n      & n          \\   	
 HD 100453            & Y                         & n                          & n               & n  & n  & n      & n          \\	     	
 HD 100546            & Y                         & Y                          & Y               & Y  & Y  & n      & Y          \\	     	
 HD 104237            & Y                         & n                          & n               & n  & Y  & Y      & n          \\	     	
 HD 135344 B          & Y                         & n                          & n               & n  & n  & n      & n          \\	     	
 HD 139614           & Y                         & n                          & n               & n  & n  & n      & n           \\	     	
 HD 141569 A         & Y                         & Y\tablefootmark{b}         & Y\tablefootmark{b} & n  & n  & n      & n        \\   	
 HD 142527           & Y                         & n                          & n               & n  & ?  & Y      & n           \\	     	
 HD 142666           & n\tablefootmark{c}        & n\tablefootmark{b}         & n\tablefootmark{b} & N  & ?  & n      & n        \\
 HD 144432           & n                         & -                          & -               & n  & n  & n      & n           \\   	
 HD 144668           & Y                         & n                          & n               & n  & n  & n      & n           \\	     	
 Oph IRS 48          & Y                         & Y                          & Y               & Y  & n  & n      & n           \\
 HD 150193           & Y                         & n\tablefootmark{b}         & n\tablefootmark{b} & n  & n  & n      & n        \\
 HD 163296           & Y                         & n                          & n               & n  & Y  & Y      & n           \\	     	
 HD 169142           & Y                         & n                          & n               & n  & n  & n      & n           \\	     	
 HD 179218           & Y                         & Y                          & Y               & n  & n  & n      & n           \\	     
 \hline                                                                                  
 DG Tau              & Y                         & Y                          & Y               & Y  & Y  & Y      & n           \\
 HT Lup              & Y                         & -                          & -               & n  & n  & n      & n           \\
 RU Lup              & Y                         & -                          & -               & n  & Y  & n      & n           \\
 RY Lup              & Y                         & -                          & -               & n  & n  & n      & n           \\
 AS 205              & Y                         & Y                          & n               & Y  & Y  & Y      & n           \\
 EM* SR 21           & n                         & n                          & Y               & n  & n  & n      & n           \\
 RNO 90              & Y                         & -                          & -               & n  & Y  & Y      & n           \\
 S Cra               & Y                         & Y                          & n               & Y  & Y  & Y      & n           \\
\hline\hline
\end{tabular}
}
\tablefoot{
\tablefoottext{a} The analysis of the CO lines is presented in paper II. The symbol ``-'' means species not observed.\\
\tablefoottext{b} Data not available in DIGIT. Line observed by \citet{Meeus12}. \\
\tablefoottext{c} Line detected by \citet{Meeus12}. 
}
\end{table}

\begin{table}
\caption{[\ion{O}{i}] and [\ion{C}{ii}] line fluxes}
\label{tab:atomic}
\centering
\begin{tabular}{llll}
\hline\hline
           & [\ion{O}{i}] \  63\,\micron & [\ion{O}{i}] \ 145\,\micron & [\ion{C}{ii}]\tablefootmark{a}\\
\hline
AB Aur      & 94.6 $\pm$ 5.2  &  3.7 $\pm$ 0.7 &  2.0     \\
HD 35187    &  4.8 $\pm$ 2.0  & -- 	       &  --      \\        
HD 36112    &  5.6 $\pm$ 0.7  &      $<$   1.1 &  $<$ 1.2 \\
HD 38120    &  7.6 $\pm$ 0.8  &  0.7 $\pm$ 0.1 &  3.3     \\
HD 50138    &  240 $\pm$  10  &  6.6 $\pm$ 0.2 &  7.8     \\
HD 97048    &  136 $\pm$   5  &  5.3 $\pm$ 0.5 &  6.3     \\
HD 98922    & 23.1 $\pm$ 1.2  & -- 	       &  --      \\        
HD 100453   & 10.2 $\pm$ 0.7  &      $<$   1.2 &  $<$ 1.3 \\
HD 100546   &  596 $\pm$   6  & 21.1 $\pm$ 1.1 & 17.6     \\
HD 104237   &  7.4 $\pm$ 0.7  &      $<$   1.5 &  $<$ 1.5 \\
HD 135344 B &  3.6 $\pm$ 0.5  &      $<$   1.2 &  $<$ 1.4 \\
HD 139614   &  3.1 $\pm$ 0.4  &      $<$   1.2 &  $<$ 1.3 \\
HD 141569 A & 25.3 $\pm$ 1.5  & -- 	       &  --      \\        
HD 142666   &      $<$   50   & -- 	       &  --      \\        
HD 142527   &  3.6 $\pm$ 0.8  &      $<$   2.9 &  $<$ 2.8 \\
HD 144432   &      $<$   5.6  & -- 	       &  --      \\        
HD 144668   & 13.3 $\pm$ 1.0  &      $<$   0.9 &  $<$ 1.1 \\
Oph IRS 48  & 30.8 $\pm$ 1.5  &  2.9 $\pm$ 0.6 &  1.2     \\
HD 150193   &  3.2 $\pm$ 0.7  & -- 	       &  --      \\        
HD 163296   & 18.2 $\pm$ 0.9  &      $<$   1.3 &  $<$ 1.3 \\
HD 169142   &  8.9 $\pm$ 2.0  &      $<$   2.2 &  $<$ 2.5 \\
HD 179218   & 17.9 $\pm$ 0.9  & 0.95 $\pm$ 0.1 &  0.4\tablefootmark{b}     \\
\hline                                            
DG Tau      & 153  $\pm$ 2.0  & 8.3 $\pm$ 0.4  &  7.4     \\  
HT Lup      &  4.0 $\pm$ 0.8  & -- 	       &  --      \\        
RU Lup      & 18.9 $\pm$ 1.2  & -- 	       &  --      \\        
RY Lup      &  5.0 $\pm$ 2.0  & -- 	       &  --      \\        
AS 205      & 21.5 $\pm$ 1.4  & 1.6 $\pm$ 0.4  &  $<$ 1.5 \\  
EM* SR 21   &      $<$   5.4  &     $<$   1.3  &  0.13    \\  
RNO 90      & 12.5 $\pm$ 1.0  & -- 	       &  --      \\        
S Cra       & 43.6 $\pm$ 1.3  & 1.8 $\pm$ 0.5  &  $<$ 1.7 \\  
\hline\hline
\end{tabular}
\tablefoot{Flux unit 10$^{-17}$ W m$^{-2}$. Flux uncertainties refer to 1\,$\sigma$ error.
For non detection the 3\,$\sigma$ upper limit is given. 
\tablefoottext{a} After subtraction of the extended emission.
\tablefoottext{b} \cii \ emission is only detected in the central spaxel.}
\end{table}

\section{Results}\label{sec:results}
\subsection{Overview}\label{sec:overview}
An overview of the detected atomic and molecular species is shown in Table \ref{tab:overview}. 
Fig. \ref{fig:spec1} 
shows the continuum normalized PACS spectrum of a T Tauri star (AS 205) and of an
Herbig AeBe star (HD 97048). Fig.~\ref{fig:spec2} and \ref{fig:spec3} show a portion 
of the PACS spectrum (continuum normalized) of selected sources. The strongest and most common 
feature is the [\ion{O}{i}]\,63\,\micron \ line, seen in all but 4 sources. The
[\ion{O}{i}]\,145\,\micron \ and [\ion{C}{ii}]\,157\,\micron \ lines
are also detected, usually in the same sources, although the detection
rate is much lower for these two lines. 
Four molecular species are seen: CO, OH, H$_2$O and CH$^+$. 
Line fluxes are reported in Tables~\ref{tab:atomic}-\ref{tab:h2o} and \ref{tab:oh}.
The CO lines are presented in paper II.  After
[\ion{O}{i}]\,63\,\micron, OH emission is the most common feature,
detected in 40\% of the sources with full spectral coverage.

\smallskip 
\noindent
We searched for other species such as [\ion{N}{ii}], HD and OH$^+$. The HD $J$=1-0 line at 112\,\micron \ 
has been detected towards TW Hya with a flux of 6.3 ($\pm$ 0.7)\,$10^{-18}$\,W\,m$^{-2}$ after deep integration 
\citep{Bergin13}. None of the sources analyzed here shows evidence of [\ion{N}{ii}], HD or OH$^+$ emission 
with 3\,$\sigma$ upper limits of the order of $1-2$\,10$^{-17}$\,W\,m$^{-2}$ for most of the sources. Typical upper 
limits in different parts of individual PACS spectra can be derived from upper limits on nearby OH lines in Table 
\ref{tab:oh}.

\subsection{ {\rm [\ion{O}{i}]} }
The [\ion{O}{i}]\,63\,\micron \ line is the most common and strongest
feature detected throughout the whole sample. The only sources in
which the line is not detected are HD 142666, HD 144432 and
SR 21. The line flux ranges from $10^{-17} - 10^{-15}$\,W\,m$^{-2}$. 
The \oib \ line is detected in 7 (out of 16) HAeBe stars and in 3 
(out of 4) T Tauri stars.  
In both cases, the spatial distribution of the line emission 
in the PACS array follows the shape of the PSF and the emission is 
not spatially extended. Fig.~\ref{fig:spec2} and \ref{fig:spec3} 
show the [\ion{O}{i}] spectra for a selected sample. 

\smallskip
\noindent
Excess emission is detected outside the central spaxel toward DG Tau (see appendix) 
in agreement with \citet{Podio12}.  In this case, the fluxes of 
the \oia \  lines are lower from those reported by \citet{Podio12} who 
computed the line fluxes by adding all the spaxels (thus including off-source emission).
The \oia \ line flux of DG Tau in Table~\ref{tab:atomic} refers
to the on-source position only (spectrum extracted from the central spaxel and 
corrected for PSF-loss, see appendix).

\subsection{{\rm OH}}
The most common molecular species detected in the PACS spectra is the
hydroxyl radical, OH. Six OH doublets with upper energy levels up to
875\,K are found including a cross-ladder transition $^2\Pi_{1/2} \
- \ ^2\Pi_{3/2} \ J = 1/2 - 3/2$ at 79\,\micron. No spatially extended
OH emission is detected outside the central spaxel of the PACS
array. The emission is seen in both Herbig AeBe groups (flared and
flat) as well as in T Tauri stars (Fig.~\ref{fig:spec1}-\ref{fig:spec3}).

\subsection{{\rm H$_2$O}}
\begin{figure}
\centering
\includegraphics[width=1\hsize]{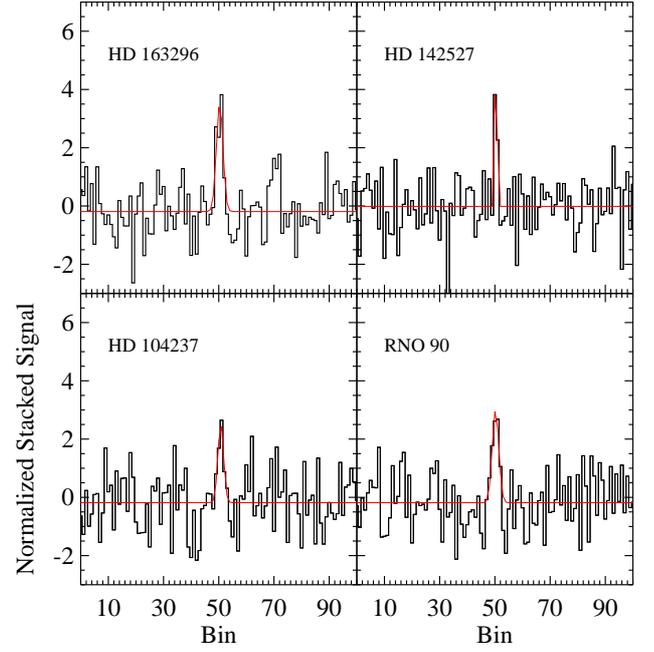}
\caption{H$_2$O line stacking for the Herbig AeBe sources HD 163296, HD 142527 and HD 104237 and for the T Tauri star
RNO 90. The stacked spectrum is divided by the standard deviation of the baseline. }
\label{fig:stack}
\end{figure}

H$_2$O lines are detected toward the T Tauri sources AS 205, DG Tau, and S CrA, including transitions from 
high-excitation levels ($E_u \sim 1000\,$K). Different transitions are detected in different targets and, 
interestingly, the strongest lines come from high energy levels in contrast to embedded sources where the 
strongest lines are from low energy levels \citep[e.g.,][]{Herczeg12}. These differences are likely due to 
different excitation mechanisms (e.g. collisions, infrared pumping, shocks) and different physical 
conditions (temperature and column density). The non detection of low-energy lines is further discussed 
in \S~\ref{sec:analysis_h2o}. The target with the richest H$_2$O spectrum is AS 205 with 10 lines detected. 
Individual line fluxes are reported in Table~\ref{tab:h2o} together with 3\,$\sigma$ upper limits to some 
of the low-energy backbone lines for AS 205. 
Far-IR H$_2$O emission in DG Tau has also been detected by \citet{Podio12} using PACS.
The line fluxes agree within 10-30\% due to different flux calibration. 
Weak H$_2$O emission is also detected toward RNO 90 through line stacking as shown in 
Fig.~\ref{fig:stack} (see below for details of the method).

\smallskip
\noindent
Herbig AeBe sources show weak or no H$_2$O far-IR emission. Weak lines have been reported
toward HD 163296 \citep{Fedele12,Meeus12} and have been confirmed
through a stacking analysis.  Two other Herbig AeBe stars show
hints of H$_2$O emission: HD 142527 and HD 104237.  
The lines are weak, with line fluxes ranging between a few $10^{-18}$\,W\,m$^{-2}$ to a
few $10^{-17}$\,W\,m$^{-2}$, often below the 3\,$\sigma$ limit.  
To confirm the presence of H$_2$O emission in these sources,
we performed a line stacking analysis as described in detail in
\citet{Fedele12}. In brief, the stacking consists in averaging the
spectral segments containing H$_2$O lines, based on a template of
observed H$_2$O lines by \citet{Herczeg12}.  Spectral bins containing
other emission lines ([\ion{O}{i}], OH, CO and CH$^+$) are masked, and
blended H$_2$O lines are excluded from the analysis. The stacked
H$_2$O spectra of HD 163296, HD 142527, HD 104237 and of the T Tauri source RNO 90
are shown in Fig. \ref{fig:stack}. 
The false alarm probability, i.e. the probability to detect a signal of equal intensity 
by stacking random portions of the PACS spectrum, is measured by counting the occurrences of 
detection in a simulation of 50,000 random stackings (after masking the the spectral bins 
containing H$_2$O, OH, CO, CH$^+$, [\ion{O}{i}] and [\ion{C}{ii}] lines). More details are given
in \citet{Fedele12}. The false alarm probability is 0.02\,\% 
for HD 142527, 0.2\,\% for HD 104237 and 0.6\,\% for RNO 90 
based on 50,000 randomized tests compared to a false
alarm probability of $<$ 0.03\,\% for HD 163296. None of the other
sources show evidence for the presence of warm H$_2$O.

\begin{figure}
\centering
\includegraphics[width=0.9\hsize]{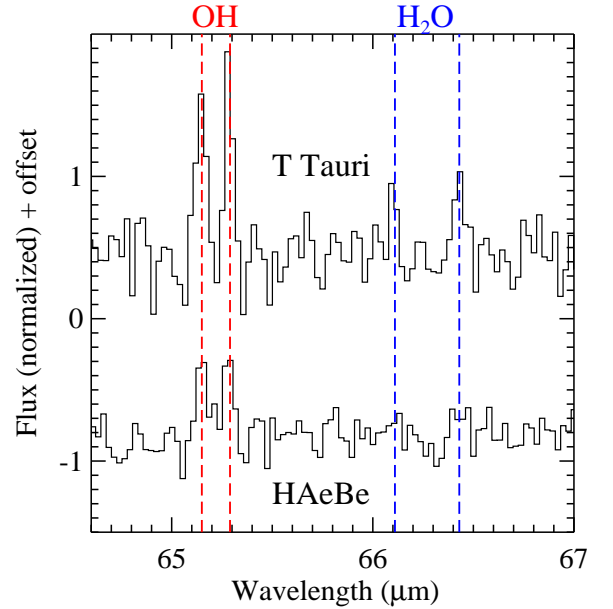}
\caption{Average PACS spectra for T Tauri and Herbig AeBe at 65\,\micron.
}
\label{fig:stack2}
\end{figure}

\smallskip
\noindent
Fig.~\ref{fig:stack2} shows the average PACS spectrum of the T Tauri and Herbig AeBe sources around 65\,\micron. 
The spectrum of each individual source is continuum subtracted and is divided by the local standard deviation. 
The source spectra in each category are then summed. The spectrum of HD 100546 was excluded from the Herbig AeBe 
list because of its lower spectral sampling. These average spectra demonstrate that OH emission is detected in 
both classes of objects, but H$_2$O only in T Tauri sources. From this result we conclude that H$_2$O far-IR 
emission is not detected in Herbig AeBe sources as a class and that the three sources with tentative detection 
through line stacking may be peculiar in this regard.  

\subsection{{\rm CH$^+$}}
CH$^+$ emission is detected toward two Herbig Ae systems: HD 100546 and HD 97048 
(Table~\ref{tab:ch+}). For HD 100546 six rotational lines are detected \citep[see also][]{Thi11} while in 
the case of HD 97048 only the $J = 6-5$ and $J = 5-4$ transitions are seen. 
The line fluxes for HD 100546 differ from those reported by \citet{Thi11} by 
10-50\% due to updated flux calibration (see \S~\ref{sec:obs_1}).

\subsection{{\rm [\ion{C}{ii}]}}\label{sec:results_CII}
\cii \ emission is detected towards 7 (out of 16) Herbig AeBe sources
and 2 (out of 4) T Tauri stars (Table ~\ref{tab:atomic}). In contrast
with [\ion{O}{i}], the \cii \ emission is often spatially extended \citep[e.g.][]{Bruderer12}. 
This proves that some of the emission
is produced in the large scale environment (cloud or remnant envelope) 
around the star even though very extended emission on $\gtrsim 6\arcmin$ 
scales has been chopped out.
More details are given in Appendix \ref{ap:cii} where the \cii \ spectral maps are also presented.
The \cii \ flux reported in Table ~\ref{tab:atomic} refers to the on-source 
spectrum only, that is the flux measured in the central spaxel after subtraction
of the spatially extended emission (see Appendix \ref{ap:cii}). These values must be considered an upper
limit to the \cii \ emission arising from the disk as extended emission from a compact
remnant envelope may still be present in the central 9\farcs4 $\times$ 9\farcs4 area of the 
sky. The closest target is at $\sim$ 100\,pc and the size of the central spaxel corresponds
to a physical scale of $\sim$ 1000\,AU which is of the same order as a compact envelope.
Moreover, given the large PSF at this wavelength, some of the spatially extended emission
will fall into the central spaxel.

\smallskip
\noindent
Two of the sources presented here (AB Aur and HD 100546) have been previously 
observed at far-IR wavelengths with ISO-LWS \citep{Giannini99, Lorenzetti02}. 
The \oia \ fluxes agree within 10 -- 15\%, which is within the 
calibration uncertainty. For the \oib \ line, the ISO flux is 1.5 times larger than 
the PACS value reported here. The \cii \ fluxes are discrepant: in both cases, the 
flux measured with ISO is much larger (more than an order of magnitude) than the 
values reported here. This is due to the diffuse \cii \ emission in the large 
(80\arcsec) ISO beam which was not removed in the ISO observations. 

\section{Analysis}

\subsection{Correlation of line luminosities}
The lines and continuum fluxes can show a correlation if the emitting conditions are physically linked. 
In particular, the emission of oxygen fine structure lines is expected to be correlated. We excluded 
the \cii \ line from this analysis as the on-source flux (i.e. the flux emerging from the disk) is only
an upper limit (see \S~\ref{sec:results_CII}).

\begin{figure*}
\centering
\includegraphics[width=0.4\hsize]{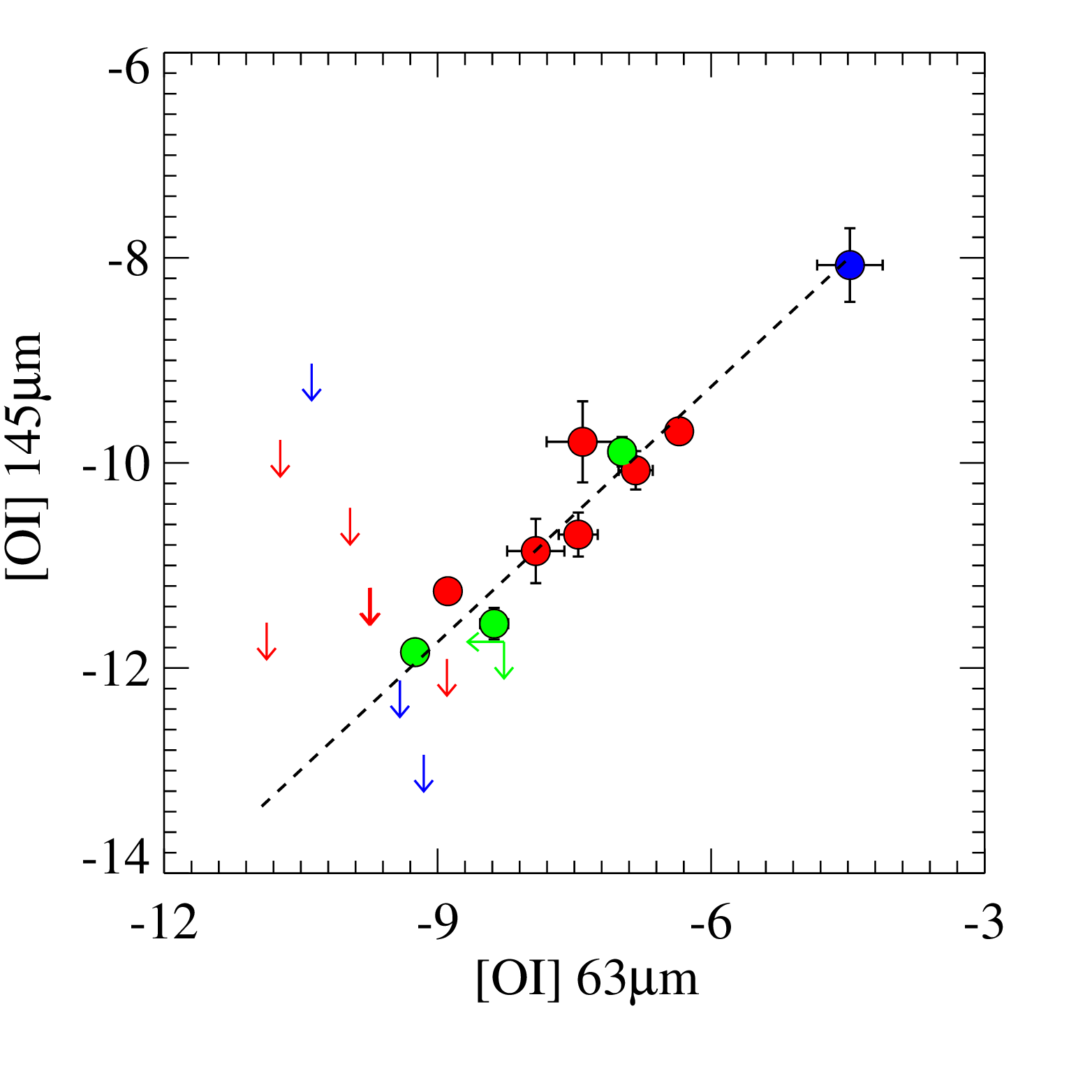}
\includegraphics[width=0.4\hsize]{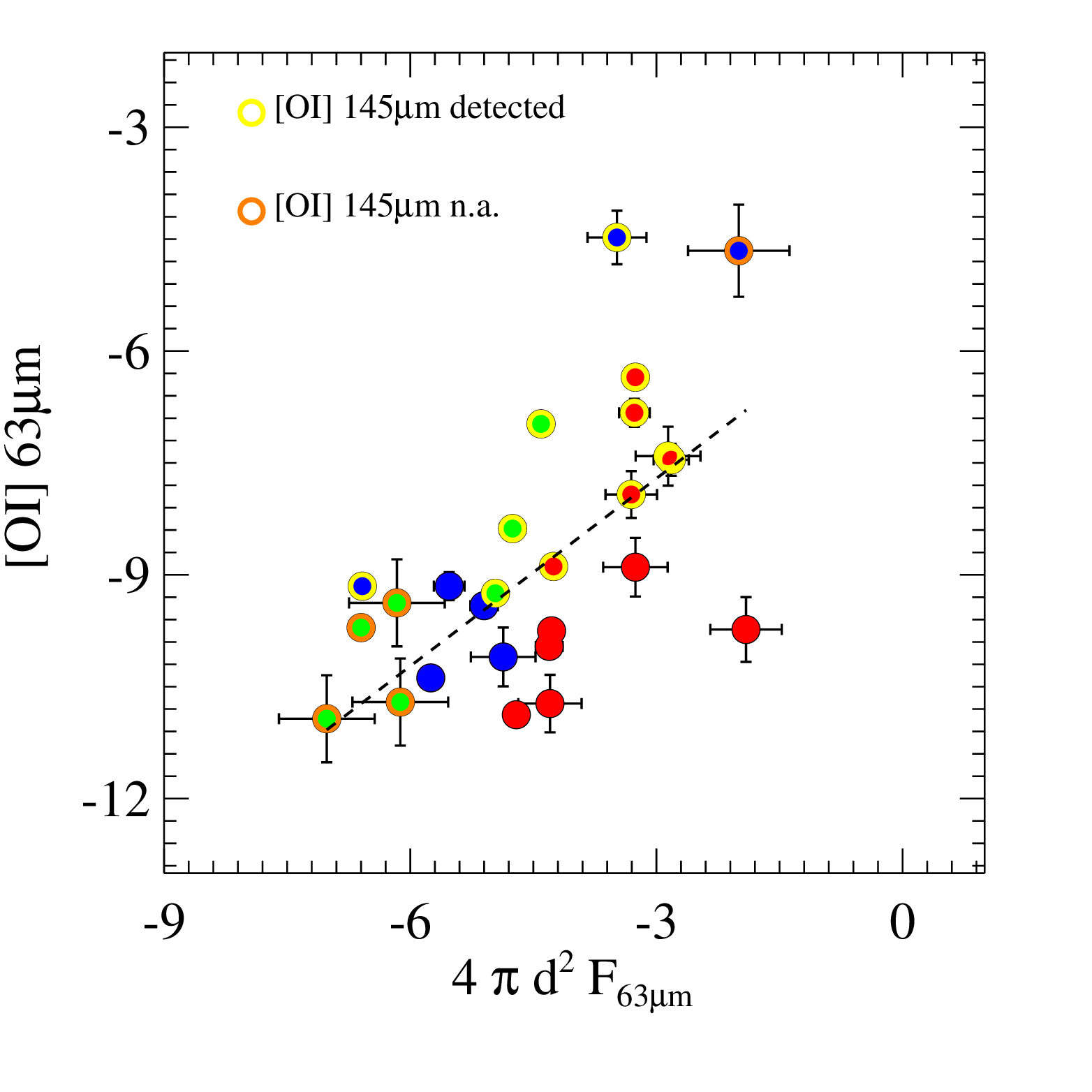}
\caption{Correlations plots. In the right panel $F_{63\,\mu {\rm m}}$ is the continuum flux at 
63\,\micron \ : open yellow circles indicate objects with \oib \ detections; open orange circles 
indicate objects with \oib \ data not available. Detections are plotted as filled circle and 
upper limits as arrows, red for HAeBe group I, blue for HAeBe group II, green for TTs.
All luminosities are expressed in $L_{\odot}$ and are plotted on a logarithmic scale.
}

\label{fig:correlations}
\end{figure*}

\smallskip
\noindent   
Fig.~\ref{fig:correlations} presents a series of plots of observed line luminosities versus each other
and versus far-IR continuum. The plotted quantities are the logarithm of line luminosity 
(log $(4 \pi d^2 F_{\rm line} / L_{\odot})$) and continuum luminosity at 63\,\micron \ 
(log $(4 \pi d^2 F_{\rm 63\,micron} / L_{\odot})$). To search for possible correlations/trends, different 
statistical tests have been performed using the ASURV \citep[Rev. 1.2][]{Isobe90, Lavalley92} statistical 
package which implements the methods presented in \citet{Isobe86}. In particular three different correlation 
tests have been used: Cox-Hazard regression, generalised $\tau$ Kendall, generalised $\rho$ Spearman. 
Linear regression coefficients are calculated with the EM algorithm. These statistical tests include upper limits.

\smallskip
\noindent
As expected, a correlation is found between the \oib \ and \oia \ luminosities  

\begin{equation}
{\rm log~L_{[\ion{O}{i}] \,145\,\mu m} = (0.83 \pm 0.06) \ log~L_{[\ion{O}{i}] \,63\,\mu m} - (4.28 \pm 0.47) }
\end{equation}

\noindent
The standard deviation is 0.28. The three correlation tests give a probability that a correlation is not
present of $<0.0002$. 

%
%

\smallskip
\noindent
We also searched for correlations between line and continuum flux. 
The only finding is that sources with stronger 
infrared continuum luminosity tend to have stronger \oia \ line luminosity (bottom panel)

\begin{equation}
{\rm log~L_{[\ion{O}{i}] 63\,\mu m} = (0.84 \pm 0.20) \ log~L_{63\,\mu m} - (5.19 \pm 0.95) }
\end{equation}
 
\noindent
with a standard deviation of 1.45. The three correlation tests give a probability of $<0.002$, suggesting
that a correlation is indeed present. Nevertheless, the scatter is large: a high infrared continuum flux
is a necessary but not sufficient condition to have stronger \oia \ emission.   
No other clear correlations with source parameters are found. The origin of these correlations and the
implications for the line emitting region are discussed in \S~\ref{sec:discussion_geometry}.

\subsection{{\rm [\ion{C}{ii}]-[\ion{O}{i}]} diagnostic plot}\label{sec:pdr}
The atomic fine structure lines can be used as diagnostics of the
physical conditions of the emitting gas.  In this section we analyze
the three line ratios: \oib \ / \oia, \oib \ / \cii, \oia \ / \cii.
The observed \oia/\oib \ ratio goes from 10--40 and it is higher than the typical ratio 
measured in molecular clouds \citep[$< 10$, e.g.,][]{Liseau99}.
The gas density and the incident FUV flux can be estimated by comparing the observations 
with PDR models.  

\noindent
In the high density regime ($n > 10^4$\,cm$^{-3}$) different PDR models do not
agree and may predict very different gas temperatures
\citep[e.g.][]{Roellig07}. Since the oxygen fine structure lines are
very sensitive to the temperature, different models produce very
different line ratios.  The aim of our analysis is to look for a trend
consistent with the observations. For this reason, the comparison of the data to a
single PDR model is justified. The model used here is from \citet{Kaufman99}. 
With this choice we can directly compare our results with those of \citet{Lorenzetti02}
based on ISO data.

\begin{figure}
\centering
\includegraphics[width=0.85\hsize]{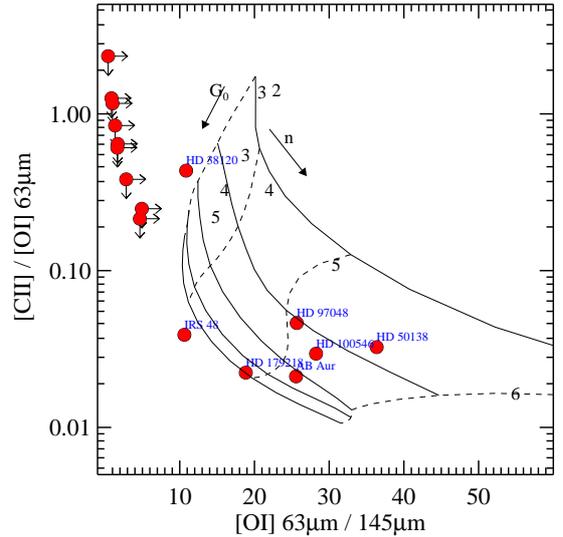}
\caption{Observed line ratios of the atomic fine structure lines and PDR model predictions.
The arrows indicate the 3-$\sigma$ upper limits.  
The continuous lines indicate the region of constant $G_0$ for values 10$^2$ - 3.6 $\times$ 10$^6$. The dashed lines indicate the iso-density surface for values of 10$^2$ - 10$^6$\,cm$^{-3}$.}
\label{fig:pdr}
\end{figure}

\noindent
Fig. \ref{fig:pdr} shows the observed line ratios and the model predictions.
DG Tau was not included in this analysis as both the \oia \ and \cii \ lines are
spatially extended and the on-source flux emission is an upper limit in both cases.
According to this model, there is a group of sources (AB
Aur, HD 50138, HD 97048, HD 100546, HD 179218) with gas density $n >$
$10^5$\,cm$^{-3}$ and $G_0$ between $10^3$ and $10^6$, where $G_0$ is the FUV 
(6 -- 13.6 eV) incident flux measured in units of the local galactic 
interstellar field \citep[1 $G_0$ = 1.6 \ 10$^{-3}$ erg\,cm$^{-2}$\,s$^{-1}$,][]{Habing68}.
These values correspond to surface temperatures $T_S \sim$ 500\,K - a few 10$^3$\,K
at radii where most of the emission originates.  The density is lower
for IRS 48 ($\sim 10^4$\,cm$^{-3}$) and HD 38120 (a few
10$^2$\,cm$^{-3}$).  As noted before, not all the \cii \ emission
measured with PACS comes from the same region as the oxygen lines,
thus the intrinsic (disk) oxygen-carbon line ratio can be higher than what
is found here.  A lower [\ion{C}{ii}]/[\ion{O}{i}] ratio shifts the
results to even higher gas density and temperature. For this reason the gas 
densities found in Fig. \ref{fig:pdr} should be considered as a lower limit
to the gas density of the oxygen emitting region. 
The values of $n$ and $G_0$ found here are larger than those found with ISO for
Herbig AeBe stars \citep{Lorenzetti02}. The differences are driven by the higher \cii \ flux
measured with ISO (see \S~\ref{sec:results_CII}). 
In general, the physical conditions derived here are consistent with disk surface layers.

\subsection{{\rm OH}, {\rm H$_2$O} and {\rm CH$^+$} excitation}
In this section the rotational diagrams of OH, H$_2$O and CH$^+$ are analyzed. 
The measured {\it Herschel}-PACS line fluxes of all sources are fit in a homogeneous way 
with a uniform slab of gas in local thermal equilibrium (LTE) including the effects of line 
opacity and line overlap \citep{Bruderer10}. 
This is a simple model to provide estimates of the physical conditions in the
regions where the lines arise. The gas column density derived here corresponds to the column density of 
a ``warm'' molecular layer. 

\subsubsection{Slab model}
The molecular emission is assumed to emerge from a disk with homogeneous 
temperature and column density and a radius $r$. The solid angle is taken to be
$d\Omega_{\rm s} = \pi r^2/d^2$, where $d$ is the distance of the source.  
The flux of an optically thin line can be written as

\begin{equation} \label{eq:rotbas}
F_{ul} = d\Omega_{\rm s} \cdot I_{ul} = \pi \frac{r^2}{d^2} \frac{h\nu_{ul}}{4\pi} A_{ul} \, N_{\rm mol} \frac{g_u e^{-E_u/k T}}{Q(T)} \,
\end{equation}

\noindent
with the line frequency $\nu_{ul}$, the Einstein-A coefficient $A_{ul}$, the 
molecular column density $N_{\rm mol}$, the statistical weight of the upper 
level $g_u$, the energy of the upper level $E_u$ and the partition function $Q(T)$.
The molecular data are from the LAMDA database \citep{Schoier05}. 
The number of emitting molecules is

\begin{equation}\label{eq:nummol}
{\mathcal N} = \frac{4 \ \pi \ d^2 \ F_{ul} \ Q(T) \ {\rm exp}(E_u/k T)}{h \ \nu_{ul} \  A_{ul} \ g_u}
\end{equation}

\noindent
Rearranging eq. \ref{eq:rotbas} yields

\begin{equation}
e^Y \equiv \frac{4 \pi F_{ul}}{A_{ul} h\nu_{ul} g_u} =  \pi \frac{r^2}{d^2} N_{\rm mol} \frac{e^{-E_u/k T}}{Q(T)} \equiv \pi \frac{r^2}{d^2} \frac{N_u}{g_u} \
\end{equation}

\noindent
Thus the vertical axis of a rotational diagram is given by

\begin{equation}
Y= \ln\left(\frac{4 \pi F_{ul}}{A_{ul} h\nu_{ul} g_u}\right) = \ln\left(  \pi \frac{r^2}{d^2} \frac{N_{\rm mol}}{Q(T)}\right) - \frac{E_u}{k T}  
\end{equation}

\noindent
The free parameters of the model are the excitation temperature \texi
\ and the column density \ncol. The emitting area can be determined
uniquely for every given combination of \texi \ and \ncol. If all 
lines are optically thin, the column density and emitting area ($\pi r^2$)
are degenerate. In this case we can measure the total number of
molecules ($\mathcal{N}$) and constrain the upper limit of \ncol \ and the lower
limit of $r$. For optically thick lines, the spectrum is calculated on a very fine wavelength grid using

\begin{equation}
I_{\nu} = d\Omega_s B_{\nu}(T_{ex}) (1-e^{\tau_{\nu}})
\end{equation}

\noindent
with $\tau_{\nu}$ obtained from the sum of the

\begin{equation}
\tau^i_{\nu} = \frac{A_{ul} c^2}{8 \pi \nu^2} (N_l \frac{g_u}{g_l} - N_u) \phi_{\nu}
\end{equation}

\noindent
over all fine structure components ($i=1, 2, \ldots$). Here, $\phi_{\nu}$ is the 
normalized line profile function, which is assumed to be a Gaussian with width 
corresponding to the thermal line width. No further (e.g. turbulent) line 
broadening is included. More details are given in \citet{Bruderer10}.
For the analysis of the H$_2$O lines an ortho-to-para ratio of 3 is assumed.
The best fit parameters are found by minimizing the reduced $\chi^2$ (\redchis) between model and observations.

\subsubsection{{\rm OH}}
OH rotational diagrams have been fitted only for sources for which 4
(or more) OH doublets have been detected. The OH rotational diagrams
are presented in Fig. \ref{fig:oh-rotdiag}-\ref{fig:oh-rotdiag2} where the 
PACS measurements are shown as red dots and the best-fit model as blue stars. 
The figure also shows the \redchis \ contours of the fit to the data; that of HD
163296 is reported in \citet{Fedele12}. The blue contour represents the 1\,$\sigma$
confidence level of the fit which corresponds to \redchis \ = minimum(\redchis) + 1.
The best fit results are reported in 
Table \ref{tab:slab}.
The OH emission is characterized by a warm temperature with \texi \ $\sim 100-400\,$K. 
In some cases all the OH lines are optically thin
(\ncol \ $\lesssim 10^{15}$\,\column) and they fall on a straight line in the 
corresponding rotational diagram.
For these sources, the OH column density and
emitting radius are degenerate so only a lower boundary to the emitting radius is given, 
varying between 20 and 50\,AU. The lowest excitation temperature is found for HD 50138 and 
DG Tau (\texi \ $\sim$ 100-130\,K). 

\smallskip
\noindent
Given the large critical density of the far-IR OH lines and the strong infrared continuum, 
non-LTE excitation (including infrared pumping) can be important. We verified the effects of non-LTE
excitation using RADEX \citep{vdTak07}. The detailed analysis is presented in Appendix~\ref{ap:nlte}.
The RADEX simulation shows that high gas densities ($n \geq 10^{10}\,$cm$^{-3}$) are needed to reproduce
the observed rotational diagram, even when a realistic infrared radiation field produced
by the dust continuum is included in the RADEX simulation. The high density justifies the LTE assumption.

\subsubsection{{\rm H$_2$O}}\label{sec:analysis_h2o} 
Fig.~\ref{fig:h2o} (right) shows the \redchis \ contours for AS 205, best fit results are given in 
Table~\ref{tab:slab}. The molecular data are listed in Table~\ref{tab:moldata}. The best fit models 
(first \redchis \ contour) give \texi \ $\sim 100-300$\,K, a column density \ncol \ $> 10^{17}$\,\column \ 
and an emitting radius $r \sim 10-30$\,AU. The rotational diagram is shown in Fig.~\ref{fig:h2o} 
(left) together with the model predictions. The model reproduces well the measured line fluxes 
(${\tilde \chi^2} = 0.5$) as well as the upper limits of the low-energy backbone lines, $3_{03}-2_{12} \ 
(E_{\rm u} = 196\,{\rm K})$ and $2_{12}-1_{01} \ (E_{\rm u}) = 114\,{\rm K}$. According to the slab model, 
the detected lines are optically thick with optical depth $\tau \sim 1 - 10$, so the inferred number 
of molecules is a lower limit. In the case of LTE, the line flux ratio of the low and high energy 
lines decreases quickly with increasing temperature. To test the validity of the LTE assumption, we 
checked the line flux ratio of a low-energy transition $2_{12}-1_{01}$ (179\,\micron \ non detected) 
versus a high-energy one $7_{07}-6_{16}$ (72\,\micron). The observed ratio is $< 0.5$ ($3\,\sigma$ upper limit). 
According to the LTE model, this ratio drops below 0.5 for $T > 65\,$K and $N > 10^{17}\,$\column. Low gas 
temperatures ($T < 65\,$K) are ruled out by the detection of high-energy transitions. Large column densities 
($N_{\rm H_2O} > 10^{17}$\,\column) are needed to reproduce the observed scatter (deviation from optically thin) in 
the rotational diagram.

\smallskip
\noindent
Non-LTE excitation may also be important because of the large critical densities of the H$_2$O 
lines detected here ($n_{\rm crit} = $ a few $10^9$\,cm$^{-3}$ for $T = 300\,$K), and the scatter of the H$_2$O lines 
in the rotational diagram can also be produced by sub-thermal excitation. If this is the case, the kinetic 
temperature of the H$_2$O containing gas is likely larger than the excitation temperature \citep[e.g.,][]{Herczeg12}. 
However, given the results of the OH modeling, the far-IR H$_2$O emission likely comes from a high gas density
region ($n \geq 10^{10}\,$cm$^{-3}$) where the H$_2$O rotational levels are in LTE.

\noindent
In the case of DG Tau, S CrA and RNO 90 only a few H$_2$O lines are detected and 
the  fit is not constrained.  

\noindent
The analysis of HD 163296 is reported in \citet{Fedele12}, who find
that the far-IR H$_2$O emission is optically thin, \ncol \
$\lesssim 10^{15}$\,\column \ with an emitting radius $r \sim
20$\,AU, and the excitation temperature is \texi $\sim 200-300$\,K (Table \ref{tab:slab}). 
For the other Herbig AeBe sources, HD 142527 and HD 104237, the individual H$_2$O lines are
too weak (below 3\,$\sigma$) for such an analysis. With the assumption of optically thin emission, 
the upper limit to the number of warm water molecules is given by Eq.~\ref{eq:nummol}. 
The $7_{07}-6_{16}$ line at 72\,\micron \ is used. The partition sum is taken from the HITRAN TIPS 
program \citep{Laraia11}
The typical upper limit to the total number of warm water molecules ranges from $\sim$ a few $10^{43} - 10^{45}$ for 
$T = 300\,$K. This number decreases by $\sim$ 20\% for a gas temperature $T = 400\,$K.
Assuming a characteristic emitting radius of 30\,AU the upper limit to the water
column density, $N_{\rm mol} = {\mathcal N}({\rm H_2O})/(\pi r^2)$, is a few 10$^{14}\,$cm$^{-2}$ 
(10$^{15}\,$cm$^{-2}$ for HD 50138).

\begin{figure*}[!ht]
\centering
\includegraphics[bb=56 0 622 1189, width=0.5\hsize]{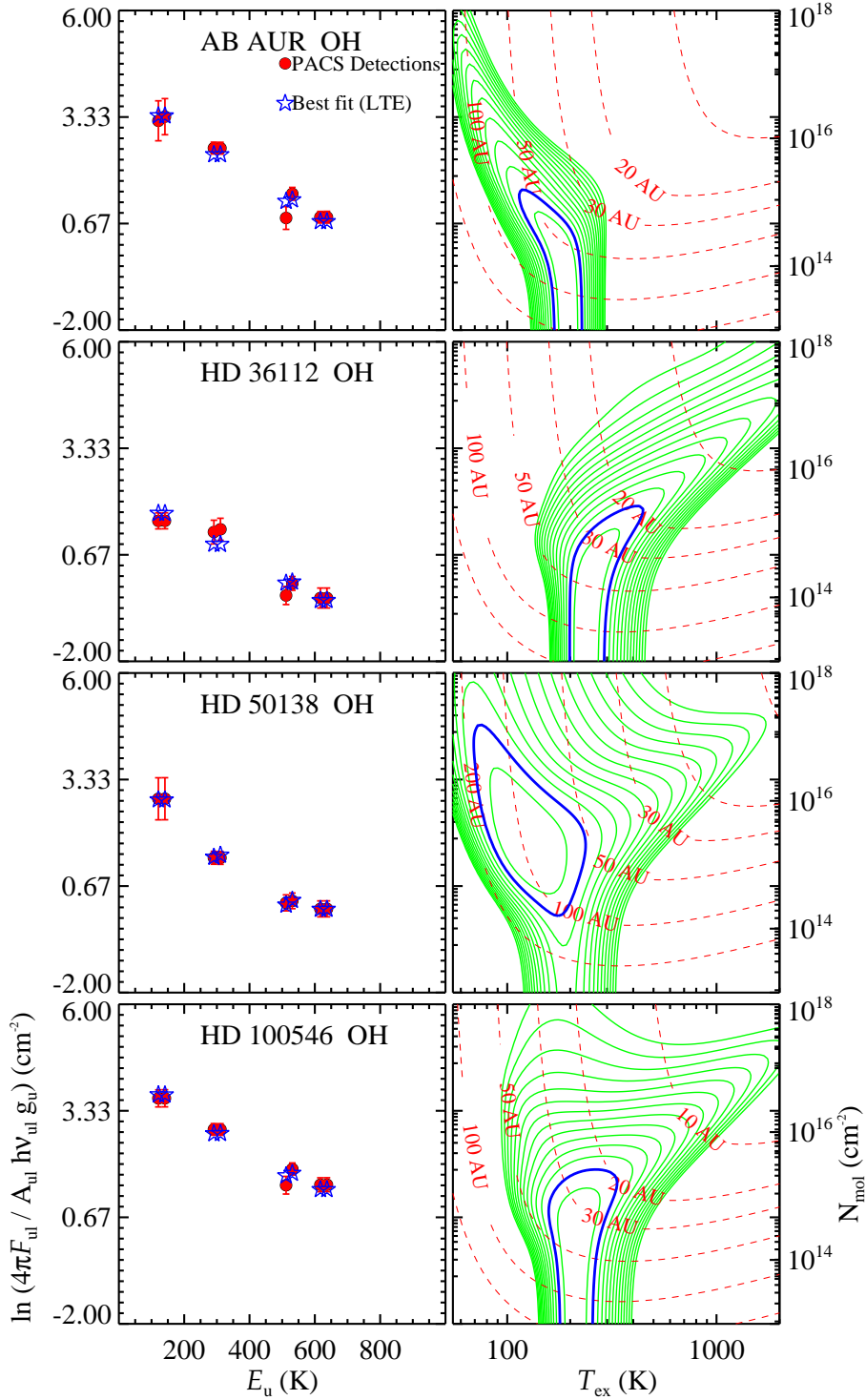}
\caption{(left) OH rotational diagram. PACS detections are plotted as (red) dots, best-fit model is shown as (blue) 
stars. (right) \redchis \ contours. The 1\,$\sigma$ confidence level is highlighted by a (blue) thick line.
The (red) dashed lines represent the emitting radius. The green lines are the \redchis \ contours in step of 0.5. }
\label{fig:oh-rotdiag}
\end{figure*}

\addtocounter{figure}{-1}
\begin{figure*}[!ht]
\centering
\includegraphics[bb=56 250 622 1189, width=0.5\hsize]{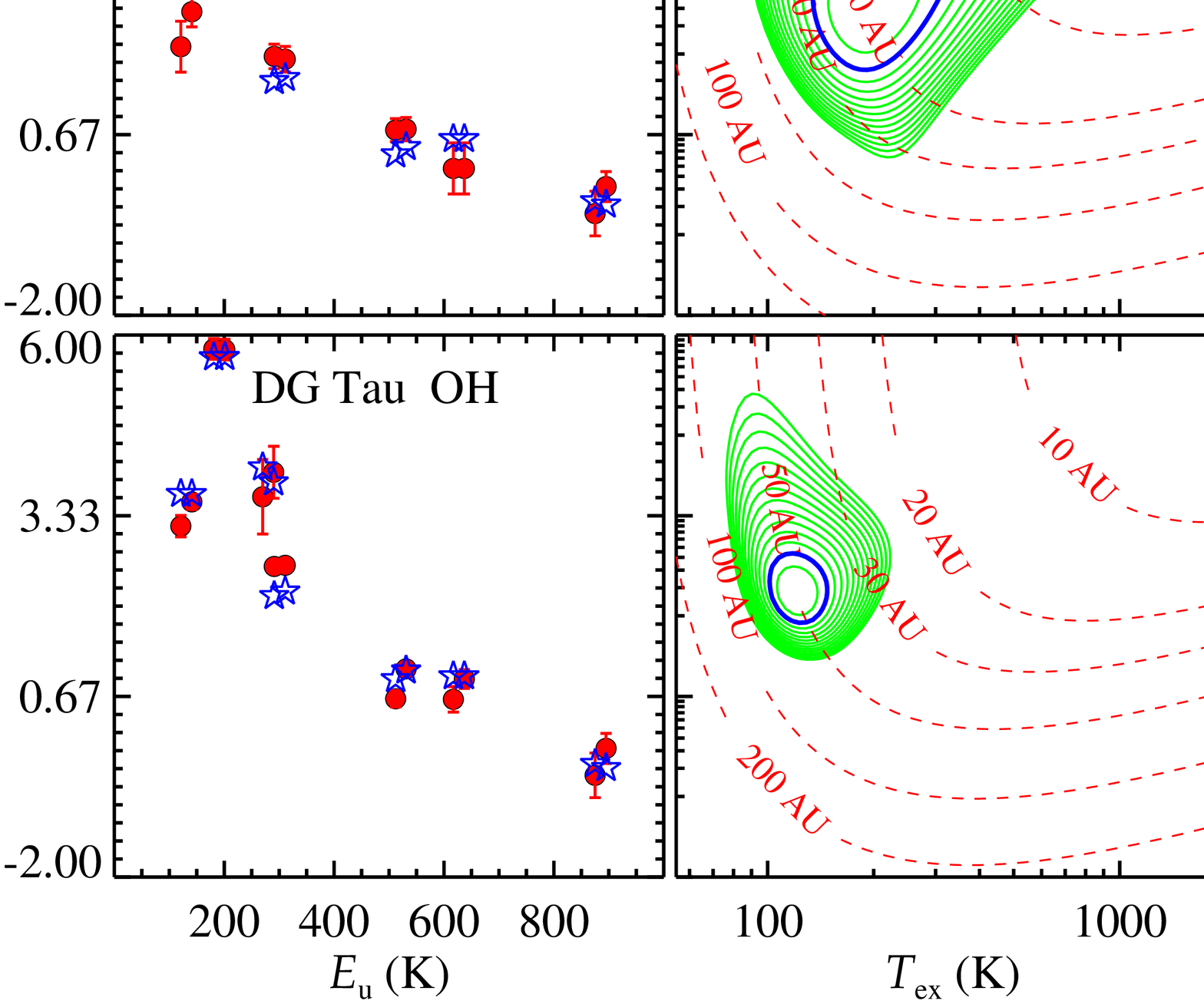}
\caption{Continued}
\label{fig:oh-rotdiag2}
\end{figure*}

\subsubsection{CH$^+$}
The CH$^+$ rotational diagram for HD 100546 is shown in Fig.~\ref{fig:ch+}. The model that best fits the 
data gives \texi \ $\sim \,80 - 120$\,K, \ncol \ $\sim 10^{16} - 10^{17}$\,\column \ and $r \sim 50 - 70\,$AU. The 
column density is not well constrained because of the large uncertainty in the two lower $J$ lines. 

\smallskip
\noindent
Compared to \citet{Thi11}, the slab model analysis presented here indicates a lower excitation temperature, 
\texi \ $\sim 100\,$K versus $323^{+2320}_{-151}\,$K \citep{Thi11}.
The CH$^+$ emitting region is also different: \citet{Thi11} find that most of the emission comes from a narrow
rim at the cavity edge between 10-13\,AU from the star through full thermo-chemical modeling, whereas the slab 
model suggests a much larger emitting area. Part of the discrepancy is due to a different flux calibration of 
the PACS spectra; the PACS spectrum presented here matches the PACS photometric points which have an accuracy 
of 5\% in absolute flux.

\begin{figure*}
\centering
\includegraphics[bb=56 850 622 1189, width=0.5\hsize]{fedele2013_f8.ps}
\caption{H$_2$O rotational diagram (left) and \redchis \ contours (right)
  for AS 205. Colours and symbols as in Fig. \ref{fig:oh-rotdiag}. The (green) arrows indicate the 3\,$\sigma$
  upper limits. 
}\label{fig:h2o}
\end{figure*}

\begin{figure*}
\centering
\includegraphics[bb=56 850 622 1189, width=0.5\hsize]{fedele2013_f9.ps}
\caption{CH$^+$ rotational diagram (left) and \redchis \ contours (right)
  for HD 100546.  Colours and symbols as in Fig. \ref{fig:oh-rotdiag}.}\label{fig:ch+}
\end{figure*}

\begin{table}
\caption{CH$^+$ line fluxes}
\label{tab:ch+}
\centering
\begin{tabular}{llll}
\hline\hline
Transition & Wavelength & HD 100546 & HD 97048\\
           & (\micron)  &     &      \\
\hline
$J=6-5$  &  60.25  & 18.5 $\pm$ 2.0 & 2.9 $\pm$ 1.5  \\
$J=5-4$  &  72.14  & 14.8 $\pm$ 2.0 & 2.2 $\pm$ 0.5  \\
$J=4-3$  &  90.02  & 13.1 $\pm$ 2.0 & $<$ 3.0 \\
$J=3-2$  &  119.86 &  3.6 $\pm$ 1.5 & $<$ 2.5 \\
$J=2-1$  &  179.60 &  4.2 $\pm$ 1.5 & $<$ 2.7 \\    
\hline\hline
\end{tabular}
\tablefoot{Units and upper limits as in Table \ref{tab:atomic}.}
\end{table}

\begin{table}
\caption{H$_2$O line fluxes.}
\label{tab:h2o}
\centering
\begin{tabular}{lll}
\hline\hline
Transition & Wavelength & AS 205 \\
           & (\micron) &   \\
\hline
$4_{32}-3_{21}$ & 58.71 & 2.5 $\pm$  1.1 \\
$7_{26}-6_{15}$ & 59.99 & 2.4 $\pm$  0.7 \\
$7_{16}-6_{25}$ & 66.09 & 3.0 $\pm$  1.1 \\
$3_{30}-2_{21}$ & 66.44 & 2.9 $\pm$  1.1 \\
$7_{07}-6_{16}$ & 71.96 & 3.1 $\pm$  1.1 \\
$3_{21}-2_{12}$ & 75.39 & 2.0 $\pm$  0.8 \\
$4_{23}-3_{12}$ & 78.74 & 2.7 $\pm$  1.0 \\
$6_{15}-5_{24}$ & 78.93 & 2.8 $\pm$  1.0 \\
$6_{06}-5_{15}$ & 83.29 & 1.7 $\pm$  0.7 \\
$2_{21}-1_{10}$ & 108.07 & 1.0 $\pm$ 0.4 \\
$3_{03}-2_{12}$ & 174.62 & $<$ 1.6 \\
$2_{11}-1_{01}$ & 179.53 & $<$ 1.6 \\
\hline
&  & DG Tau\\
\hline
$4_{23}-3_{12}$ &  78.75 & 1.2 $\pm$ 0.6 \\
$6_{16}-5_{05}$ &  82.03 & 1.3 $\pm$ 0.7 \\
$2_{21}-1_{10}$ & 108.13 & 1.3 $\pm$ 0.4 \\
$2_{12}-1_{01}$ & 179.54 & 1.1 $\pm$ 0.4 \\
\hline
&  & S CrA\\
\hline
$8_{18}-7_{07}$ &  63.31 & 2.3 $\pm$  0.6 \\    
$7_{16}-6_{25}$ &  66.09 & 1.6 $\pm$  0.6 \\    
$7_{07}-6_{16}$ &  71.96 & 2.4 $\pm$  0.7 \\    
$4_{23}-3_{12}$ &  78.74 & 2.3 $\pm$  0.7 \\    
$4_{13}-4_{04}$ & 187.11 & 1.2 $\pm$  0.4 \\    
\hline\hline
\end{tabular}
\tablefoot{Units and upper limits as in Table \ref{tab:atomic}.}
\end{table}

\begin{table}
\caption{Best fit results of the slab model}
\label{tab:slab}
\centering
\begin{tabular}{lllll}
\hline\hline
 & $T_{\rm ex}$  & $N_{\rm mol}$  & $r$   & log(${\mathcal N}$)   \\
 &  [K]        & [cm$^{-2}$]  & [AU]   &  \\
\hline
 & \multicolumn{3}{c}{OH}\\
\hline
AS  205    & 190 &  8 $\times$ 10$^{15}$ &   19    & 45.31 \\
DG  Tau    & 115 &  4 $\times$ 10$^{15}$ &   50    & 45.85 \\
AB Aur     & 190 &  $<$ 10$^{14}$        & $>$ 50  & 44.25 \\
HD  36112  & 240 &  $<$ 10$^{14}$        & $>$ 50  & 44.25 \\
HD  50138  & 130 &  2 $\times$ 10$^{15}$ &   95    & 46.10 \\
HD 100546  & 210 &  2 $\times$ 10$^{14}$ &   40    & 44.35 \\
HD 104237  & 160 &  2 $\times$ 10$^{15}$ &   20    & 44.75 \\
HD 163296  & 425 &  8 $\times$ 10$^{14}$ &   15    & 44.10 \\
\hline
 & \multicolumn{3}{c}{H$_2$O}\\
\hline
AS 205     & 100 -- 300 & $>$ 10$^{17}$      & 10 -- 30  &  $>$ 45.85 \\
HD 163296  & 250 -- 300 & $10^{14}-10^{15}$   & 20        & 43.5 - 44.5\\ 
& \multicolumn{3}{c}{CH$^+$}\\
\hline
HD 100546  &  80 -- 120 & $10^{16} - 10^{17}$ & 50 -- 70  & 46.94\\
\hline\hline
\end{tabular}
\tablefoot{Upper and lower limits are bounds from the modeling.}
\end{table}

%

\section{Discussion}

\subsection{Origin of far-IR emission lines}
According to the results of the OH rotational diagram, 
the far-IR OH lines are emitted by warm gas with \texi \ of 100 - 400\,K.
The OH emission toward AB Aur, HD 36112, HD 100546 is optically thin ($N < 10^{15}\,$cm$^{-2}$). 
For the remaining sources (AS 205, DG Tau, HD 50138, HD 163296) the far-IR OH lines 
are at the border between optically thick and optically thin emission ($N \sim 10^{15} - 10^{16}$).
The derived excitation temperature and the emitting radius are consistent with a 
disk origin with the emission coming from the upper layers of the disk at distances of 15--50\,AU
from the star (100 \,AU in the case of HD 50138). Given the high excitation temperature and high critical 
densities (of order $10^8$ cm$^{-3}$ to excite OH), the emitting radius cannot be 
much larger. 

\smallskip
\noindent
In the case of the T Tauri system DG Tau, the derived excitation temperature
is 115\,K (lowest in the sample) and the OH emitting radius is $\sim$ 50\,AU, which
is 2.5 times larger than the emitting radius of the other T Tauri star (AS 205).
In this case, a further contribution to the OH lines may come from a shock associated with the molecular 
outflow/jet associated with the system. Indeed, compact warm OH emission has been observed to be
associated with outflows in embedded young stellar objects
\citep{vanKempen10,Wampfler10,Podio12,Wampfler13,Karska13}.
Further analysis including the mid-infrared lines (from {\it Spitzer}) is needed to disentangle
the disk/outflow origin of OH for these systems. In the case of AS 205 the emitting redius of the 
far-IR H$_2$O lines is is $\sim 10-30$\,AU, also consistent with a disk origin. 

\smallskip
\noindent
As discussed in \S~\ref{sec:pdr}, the high density and UV fluxes implied
by the ratios of the atomic fine structure lines are consistent with a disk
origin for most of the sources. The \cii \ emission is spatially extended in all sources 
where the line is detected. This suggests that there is a contribution from a diffuse
cloud (or remnant envelope) around the young star. 
The on-source \cii \ flux correlates with the oxygen line fluxes suggesting that (some of) 
the on-source \cii \ emission is associated with the disk. From the PACS spectra it is however
impossible to disentangle the disk emission from the diffuse emission. HIFI spectra of the
\cii \ line profiles are needed to solve this issue (Fedele et al., in prep.).

\subsection{Disk geometry and dust settling}\label{sec:discussion_geometry}
The protoplanetary disks presented here vary in geometry and
degree of grain growth and settling.  These factors are
important for the excitation of the atomic and molecular gas. For
example, in the case of water, a combination of these factors can play
a role in the low detection rate towards Herbig AeBe systems. As
pointed out by \citet{Woitke09}, whether or not the puffed-up inner
rim shadows the hot water layer is important, since shadowing reduces
the UV radiation field by about two orders of magnitude and increases
water by the same amount. Also, grain settling, presence or absence
of PAHs and the gas-to-dust ratio can all have a large effect in
boosting line fluxes \citep[e.g.,][]{Meijerink09,Najita11,Tilling12,Bruderer12}.

\smallskip
\noindent
Far-IR CO emission is only detected in HAeBe systems of Group I (Table~\ref{tab:overview} and paper II). 
The high$-J$ CO lines detected with PACS are
sensitive to the UV flux impinging onto the disk, which controls the
disk gas temperature. Using the thermo-chemical models of
\citet{Bruderer12}, we have shown that flared disks indeed have higher gas
temperatures out to several tens of AU and stronger high$-J$ CO
fluxes (paper II). This is an independent proof that the disks of Group I indeed
have a flared geometry.  On the other hand, the OH lines are less
sensitive to the gas temperature, consistent with its detection in
both Group I and II sources (Bruderer et al., in prep. and \S~\ref{sec:chemistry}).

\smallskip
\noindent
The \oia \ line is detected in most of the disks
independently from disk geometry and stellar parameters. 
The high detection rate toward Group II sources is interesting. 
If these sources are indeed self-shadowed and/or have grain growth and
settling, the gas temperature in the surface layers should be lower
\citep{Jonkheid07} and the atomic and molecular emission at far-IR
wavelengths is expected to be reduced in Group II disks.
The excitation of the O ($^3$P$_1$) level (upper level of the
\oia \ line) is mostly due to collisions with H and H$_2$. Once the
gas density exceeds the critical density of the line\footnote{
$n_{\rm crit}$(\oia)  = 2.5$\cdot 10^5$\,\density at 100\,K and lower for higher temperature} 
the excitation
depends only on the temperature and no longer on the density. In this
scenario, Group I sources can have stronger \oia \ and \oib \ emission
due to the higher temperature of the gas.  
The intensity of the \oia \ line varies by two orders of magnitude for a given 
value of the continuum flux at 63\,\micron. This implies a different gas density 
structure (in the [\ion{O}{i}] forming region) from object to object.
According to model predictions \citep[e.g.][]{Woitke09,Bruderer12}, the oxygen 
emitting region can be more extended than the far-IR continuum. 
The FIR continuum emission comes mostly from the inner $\sim$ 50\,AU, while 
the oxygen lines originate in the outer disk (up to a few 10$^2$\,AU in the 
case of Herbig AeBe stars). The large spread in \oia \ fluxes for a given 
FIR continuum flux (Fig. \ref{fig:correlations}) suggests that the 
[\ion{O}{i}]-bright sources may have an enhanced scale-height (more flared) 
compared to the [\ion{O}{i}]-faint sources. 
Based on the results of \S~\ref{sec:pdr} and on the high detection rate of \oia \ we conclude 
that the oxygen lines have a disk origin in most of the cases, with the strength determined by
the specific disk structure.


\subsection{Comparison to near- and mid-IR spectroscopy}
The comparison of the far-IR spectra shown here to the near-
(1-5\,\micron) and mid- (10-40\,\micron) IR spectra can give 
insights on the radial distribution of different gas species in the
upper layers of protoplanetary disks. At longer wavelengths also larger
vertical depths into the disk are probed.

\begin{figure}
\centering
\includegraphics[bb=100 650 558 915, width=1.\hsize]{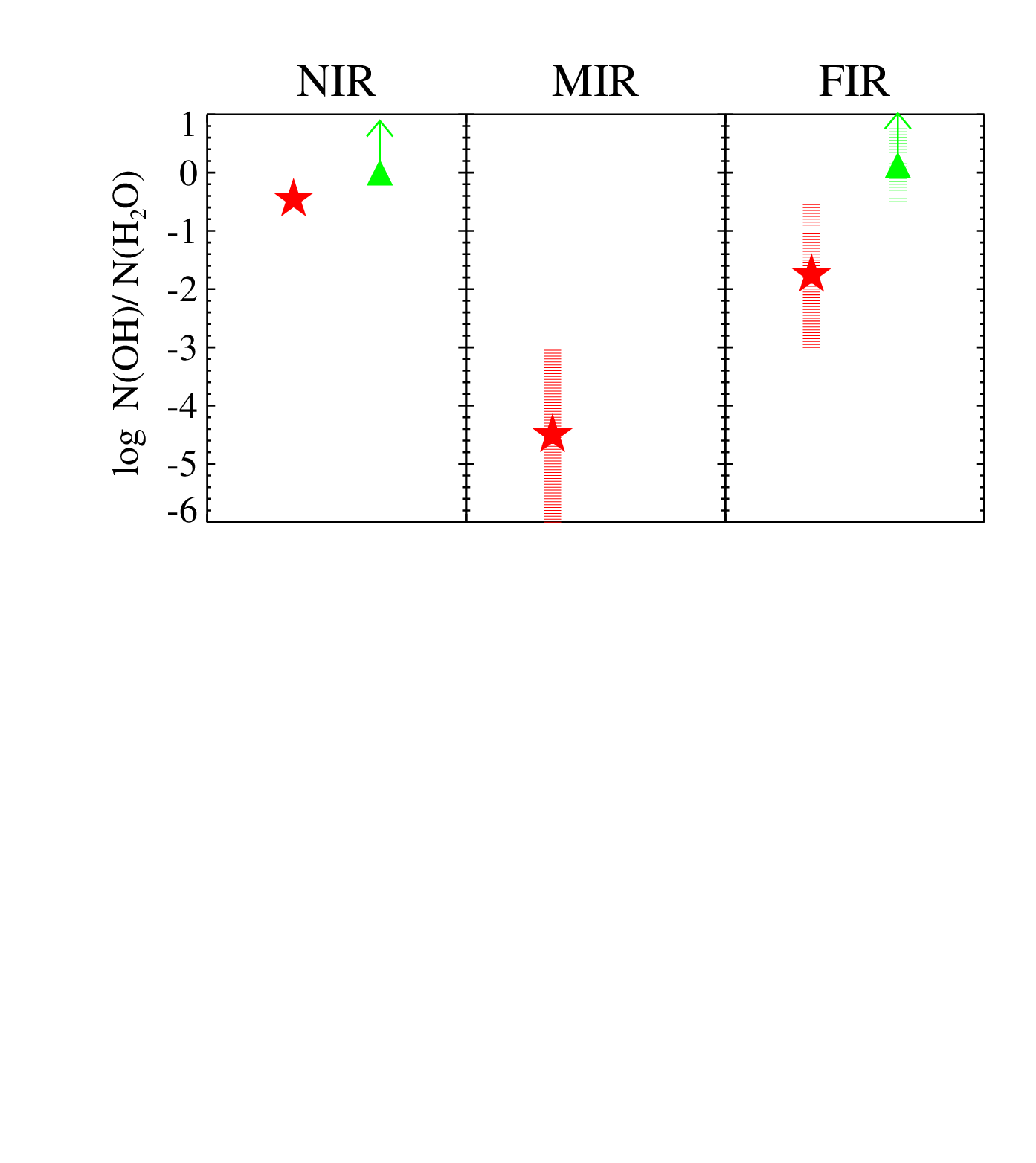}
\caption{OH/H$_2$O column density ratio for T Tauri (red stars) and Herbig AeBe (green squares) sources.
The dashed regions indicate the range of observed ratio.}
\label{fig:abundance}
\end{figure}

\smallskip
\noindent
Fig. \ref{fig:abundance} shows the OH/H$_2$O column density ratio for T Tauri and 
Herbig AeBe stars measured at different wavelengths. The values 
represent the ratio of the total number of molecules and are taken from
\citet{Salyk08}, \citet{Fedele11}, \citet{Salyk11b} and from this work. 
In the case of Herbig AeBe stars, the MIR ratio refers to the ratio of the 
upper limits and is thus not constrained.
At all wavelength ranges, the Herbig AeBe disks show a higher OH/H$_2$O
abundance ratio compared to T Tauri disks.
For Herbig AeBe systems the OH/H$_2$O lower limit does not vary much from near- 
to far-IR wavelengths. In contrast the T Tauri disks do show a clear 
decrease in the OH/H$_2$O ratio. The difference between the mid- and far-IR
ratios are not significant. 

\smallskip 
\noindent
\underline{Herbig AeBe}: The major difference between short and long wavelengths 
is the lack of any H$_2$O lines at near-IR \citep{Mandell08,Fedele11} whereas weak 
H$_2$O lines are detected at mid- and far-IR in some sources 
\citep{Pontoppidan10,Fedele12,Meeus12}. Even though the detection rate of warm H$_2$O 
is low and the individual lines are weak ($<$ a few 10$^{-17}$\,W\,m$^{-2}$), this 
finding suggests a different H$_2$O abundance between the inner and outer disk. 
However, as Fig.~\ref{fig:abundance} shows, the OH/H$_2$O abundance limits are similar 
at all wavelengths. Also interesting is the high detection rate ($\sim 40\,\%$) of 
far-IR OH emission for sources with full spectral coverage. This emission is 
detected in both Group I and II sources in contrast to near-IR OH emission which
is detected in Group I sources only \citep{Fedele11}. Finally, near-IR CO emission 
(ro-vibrational) is detected in several Herbig AeBe systems independently from disk 
geometry in contrast to far-IR CO which is only detected in flared disks (paper II). 
This difference is likely a consequence of the larger number of UV photons that impinge onto the disk 
surface in the case of a flared geometry which can heat the gas to the 
larger distances responsible for the far-IR lines.

\smallskip 
\noindent
\underline{T Tauri} : The PACS spectra presented here
are rich in molecular emission similar to the near-IR and
mid-IR spectra. AS 205 and RNO 90 have a rich OH and H$_2$O spectrum
ranging from the pure-rotational lines detected here to ro-vibrational
lines detected with NIRSPEC and CRIRES (3\,\micron) and {\it
  Spitzer}/IRS (10-40\,\micron)
\citep{Salyk11a,Pontoppidan10,Mandell12}. The energy levels involved
range from a few hundred to a few thousand K. The H$_2$O-rich PACS
spectrum is unlikely to originate from the same region of the disk as
the shorter wavelength data. Indeed, based on the slab model, the
emitting region of the far-IR lines of OH and H$_2$O has a radius $r
\sim 20-30\,$AU in contrast to a radius of a few AU (or less) for the hot H$_2$O
near-IR and mid-IR lines \citep{Salyk11a}. The
conclusion is that H$_2$O is present in the surface layers of 
disks around low-mass pre-main sequence stars from the inner ($\sim$ 0.1\,AU) 
out to the outer ($\sim$ 30\,AU) disk, but the OH/H$_2$O abundance ratio changes
with radius (see \citealt{Zhang13} for the analysis of H$_2$O data of one source).

\subsection{Disk chemistry and molecular excitation}\label{sec:chemistry}
The different detection rates and excitation mechanisms of the various
species provide information about the chemical processes governing the
atmosphere of protoplanetary disks.  In the case of the UV-bright
Herbig stars, the chemistry and excitation are regulated more strongly
by photoprocesses like UV fluorescence (e.g. for CO) and
photodissociation of molecules (e.g. OH, H$_2$O) than in T Tauri
stars, unless those low-mass sources have significant UV excess due to
accretion.

\smallskip
\noindent
First, we find that OH far-IR emission is detected in all groups of
sources in contrast to CO far-IR emission which is only detected in
flared Herbig AeBe disks and T Tauri disks. The excitation of the
high$-J$ CO rotational lines in disks is regulated by the gas
temperature which in turn is controlled by the UV radiation
field. In the case of OH, the lines are excited
either by collisions with atomic and molecular hydrogen or through
infrared pumping, both of which are much less sensitive to geometry. 
A third viable mechanism for OH is the prompt
emission after the OH molecules are produced rotationally excited by
the photodissociation of water which is also controlled by the UV field.

\smallskip
\noindent
Another interesting finding is the detection of (weak) far-IR
H$_2$O emission. The non-detection of hot H$_2$O lines at near- and
mid-IR wavelengths suggests that the atmospheres of disks around early
type stars lack H$_2$O molecules due the photodissociation
of H$_2$O by the strong UV radiation field of the central
star \citep{Fedele11, Pontoppidan10}.  In contrast, the PACS detection
of warm H$_2$O in some sources suggests that H$_2$O molecules can
survive at large distances ($>$ 30\,AU) from the star and somewhat deeper into the disk, 
likely produced
by high temperature reactions of O + H$_2$ and OH + H$_2$ driving much
of the oxygen into water \citep[e.g.,][]{Bergin11, Woitke10}. The importance of
this result is that it reveals the presence of an H$_2$O reservoir in
the outer disk around early type stars, beyond the traditional snow
line. 

\smallskip
\noindent
CH$^+$ emission is found toward HD 100546 and HD 97048. Interestingly
these are the only two Herbig AeBe sources where ro-vibrational H$_2$
emission has been detected so far \citep{Carmona11}. The velocity
profile of the line suggests extended H$_2$ emission to more that 50
\,AU (radius) from the star \citep{Carmona11}. This is likely the same
spatial region traced by the far-IR CH$^+$ lines reported here: 
the CH$^+$ emitting area in HD 100546 is 50-70\,AU according to the slab model. The
detection of vibrationally excited H$_2$ and CH$^+$ towards the same
sources is not a coincidence but relates to the gas phase reaction \citep{Sternberg95}

\begin{equation}
{\rm C^+ + H_2^* \rightarrow CH^+ + H}
\end{equation}

If H$_2^*$ is vibrationally excited, the forward reaction (which is
endothermic by $\sim$ 4000\,K) is faster \citep[see also][]{Agundez10, Thi11}.

\section{Conclusion}
We present far-IR spectra of Herbig AeBe and T Tauri stars taken with
Herschel/PACS. Besides the fine structure lines of [\ion{O}{i}] and
[\ion{C}{ii}], emission is detected of CO (paper II), OH, H$_2$O and 
CH$^+$.  The most common feature detected is the \oia \ line.

\smallskip
\noindent
far-IR OH emission is detected in several sources. An LTE slab
model including optical depth effects is used to fit the OH rotational
diagram. The OH lines are likely associated with the disk, probing a warm 
layer of gas in the outer disk ($r \gtrsim 20\,$AU).  
In contrast to the high-$J$ CO lines, the OH lines are detected in both
flat and flared disks (Group I and II) around Herbig AeBe stars. The
reason for this different may be the different excitation mechanisms for the two
species.

\smallskip
\noindent
Warm H$_2$O emission is detected in three Herbig AeBe sources and in
four T Tauri sources.  In the Herbig sources, the emission is weak and
the detection of warm H$_2$O is confirmed only by line stacking. This
result reveals the presence of an H$_2$O reservoir in the outer disk
region around Herbig stars. However, the OH/H$_2$O abundance limit is the 
same  between inner and outer disk, consistent with a decreasing UV field.
In the case of the T Tauri star AS 205 the slab model
suggests ``warm'' (\texi \ 100-300\,K) H$_2$O emission coming from the 
inner 10-30\,AU from the star. Overall, the OH/H$_2$O column density ratio
decreases from inner and outer disk for T Tauri disks.

\smallskip
\noindent
The flux ratios of atomic fine structure lines are fitted with PDR models
involving high gas density ($n > 10^5$\,cm$^{-3}$) and high UV radiation field
($G_o \sim 10^3 - 10^7$) as expected for the atmosphere of protoplanetary disks.
The presence of spatially extended \cii \ emission (on scale of 10$^3$\,AU) 
implies the presence of diffuse material (e.g., remnant of the molecular cloud) 
around the young stars.   

\begin{acknowledgements}
  Support for this work, part of the Herschel Open Time Key Project Program, was provided by NASA through an award issued 
  by the Jet Propulsion Laboratory, California Institute of Technology. We are grateful to the DIGIT team for stimulating 
  discussions and scientific support. Astrochemistry in Leiden is supported by the Netherlands Research School for Astronomy 
  (NOVA), by a Spinoza grant and grant 614.001.008 from the Netherlands Organisation for Scientific Research (NWO), and by 
  the European Community's Seventh Framework Programme FP7/2007-2013 under grant 238258 (LASSIE) and grant 291141 (CHEMPLAN). 
  PACS has been developed by a consortium of institutes led by MPE (Germany) and including UVIE (Austria); KU Leuven, CSL, IMEC
  (Belgium); CEA, LAM (France); MPIA (Germany); INAF-IFSI/OAA/OAP/OAT, LENS, SISSA (Italy); IAC (Spain).  This development has been
  supported by the funding agencies BMVIT (Austria), ESA-PRODEX (Belgium), CEA/CNES (France), DLR (Germany), ASI/INAF (Italy), and
  CICYT/MCYT (Spain). The research of M.G. has been supported by the the Austrian Research Promotion Agency (FFG) through the ASAP 
  initiative of the Austrian Federal Ministry for Transport, Innovation and Technology (BMVIT).
  We thank the anonymous referee for the helpful comments.
\end{acknowledgements}

\bibliographystyle{aa}


\newpage

\begin{appendix}

\section{PACS Spectra}
\begin{figure*}
\centering
\includegraphics[width=0.9\hsize]{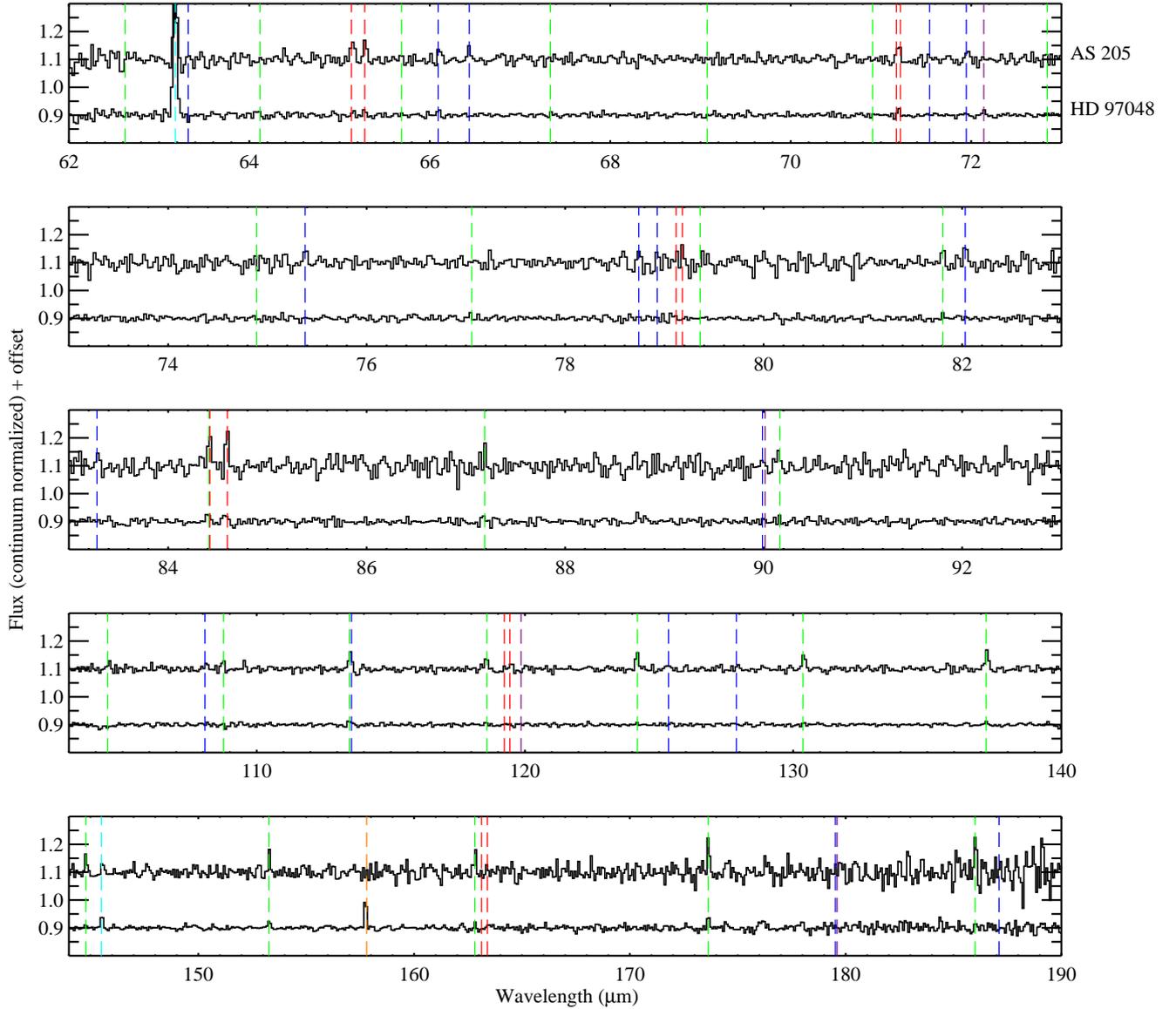}
\caption{PACS spectrum of the T Tauri star AS 205 (top) and of the Herbig Ae star HD 97048 (bottom). 
The marks indicate the positions of [\ion{O}{i}] (light blue), [\ion{C}{ii}] (orange), CO (green), 
OH (red), H$_2$O (blue) and CH$^+$ (purple) lines.}\label{fig:spec1}
\end{figure*}

\begin{figure*}
\centering
\includegraphics[width=0.9\hsize]{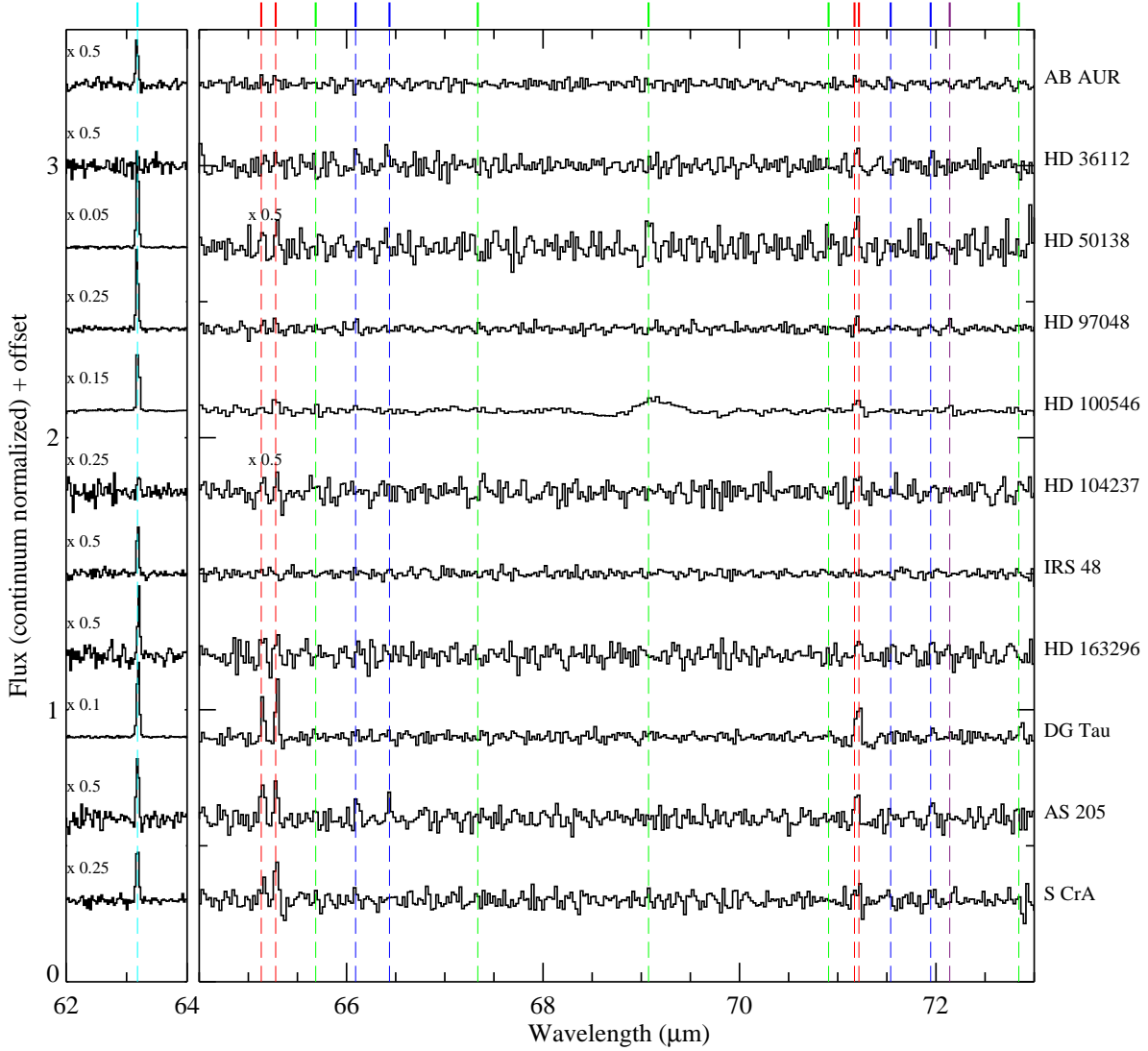}
\caption{PACS spectra of a sub-sample of the programme stars between 62-73\,\micron. Marks and colours as in 
Fig.~\ref{fig:spec1}. The 69\,\micron \ forsterite feature is present in the spectrum of 
HD 100546.}\label{fig:spec2}
\end{figure*}

\begin{figure*}
\centering
\includegraphics[width=0.9\hsize]{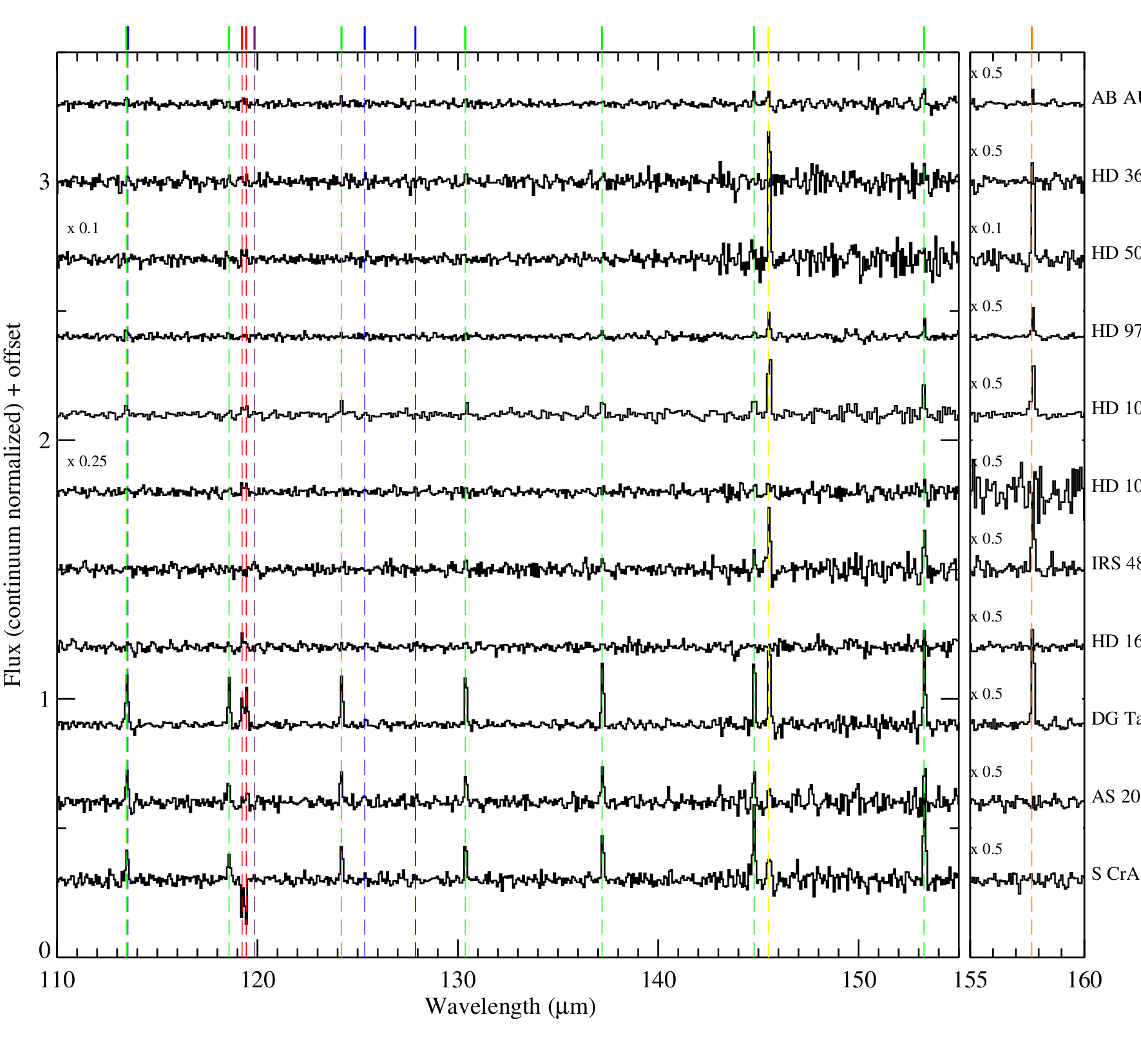}
\caption{As Fig.~\ref{fig:spec2} for the wavelength range 110-160\,\micron.}\label{fig:spec3}
\end{figure*}

Fig.~\ref{fig:spec1} shows the PACS spectra of a the T Tauri star AS 205 and of the Herbig Ae star HD 97048 between 
62--190\,\micron. The main molecular and atomic transitions detected in the whole sample are shown. 
Figs.~\ref{fig:spec2} and \ref{fig:spec3} show a portion of the PACS spectra of selected sources.

\section{OH line fluxes and molecular data of selected species}
Table~\ref{tab:oh} reports the line fluxes of the far-IR OH transitions. The line flux uncertainties correspond to the 
1\,$\sigma$ error. For non-detection the 3\,$\sigma$ upper limit is reported. Table~\ref{tab:moldata} reports the 
atomic and molecular data of the transitions detected in this paper. Molecular data are taken from the 
LAMDA database \citep{Schoier05}.

\begin{table*}
\caption{OH line fluxes}
\label{tab:oh}
\centering
\begin{tabular}{lllllllll}
\hline\hline
& \multicolumn{6}{c}{$^2\Pi_{3/2}$} &  \multicolumn{2}{c}{$^2\Pi_{1/2} - ^2\Pi_{3/2}$}\\ 
& $9/2^--7/2^+$   & $9/2^+-7/2^-$  & $7/2^--5/2^+$   & $7/2^+-5/2^-$  & $5/2^--3/2^+$    & $5/2^+-3/2^-$   & $1/2^--3/2^+$  & $1/2^+-3/2^-$\\ 
& 65.13\,\micron & 65.28\,\micron & 84.42\,\micron & 84.59\,\micron & 119.23\,\micron & 119.44\,\micron & 79.11\,\micron & 79.18\,\micron\\
\hline
AB Aur      &  7.0 $\pm$ 2.0 &  12.5 $\pm$ 2.0 &   10.1 $\pm$ 1.5    & 10.1 $\pm$ 1.5 &  2.8 $\pm$  1.4  &  3.1 $\pm$ 1.4  & $<$ 9.0 & $<$ 9.0 \\
HD 36112    &  2.2 $\pm$ 0.5 &  2.9 $\pm$  0.5 &    2.5 $\pm$ 0.8    &  2.6 $\pm$ 0.8 &  0.49 $\pm$ 0.20 & 0.52 $\pm$ 0.12 & $<$ 3.4 & $<$ 3.4 \\
HD 38120    &       $<$  3.5 &        $<$  3.5 &          $<$ 2.8    &        $<$ 2.8 &         $<$ 0.9  &        $<$ 0.9  & $<$ 2.7 & $<$ 2.7 \\
HD 50138    &  4.0 $\pm$ 0.8 &  4.1 $\pm$  0.8 & \tablefootmark{a}   &  1.9 $\pm$ 0.6 &  1.0  $\pm$ 0.2  &  1.1 $\pm$ 0.2  & $<$ 4.0 & $<$ 4.0 \\
HD 97048    &  4.9 $\pm$ 1.0 &  5.7 $\pm$  1.0 &  6.8 $\pm$ 1.0      &  6.8 $\pm$ 1.0 &         $<$ 2.4  &        $<$ 2.4  & $<$ 4.0 & $<$ 4.0 \\
HD 100453   &       $<$  3.4 &        $<$  3.4 &        $<$ 3.0      &        $<$ 3.0 &         $<$ 1.3  &        $<$ 1.3  & $<$ 2.0 & $<$ 2.0 \\
HD 100546   & 13.6 $\pm$ 0.4 & 19.9 $\pm$  3.3 & \tablefootmark{a}   & 13.8 $\pm$ 2.0 &  4.2  $\pm$ 0.9  &  4.2 $\pm$ 0.9  & $<$ 6.0 & $<$ 6.0 \\
HD 104237   &  3.0 $\pm$ 0.5 &  3.0 $\pm$  0.5 &   3.1 $\pm$ 0.5     &  3.1 $\pm$ 0.5 &  1.2  $\pm$ 0.3  &  1.2 $\pm$ 0.3  & $<$ 3.6 & $<$ 3.6 \\
HD 135344 B &       $<$  4.0 &        $<$  4.0 &         $<$ 3.8     &        $<$ 3.8 &  0.67 $\pm$ 0.19 & 0.71 $\pm$ 0.16 & $<$ 3.0 & $<$ 3.0 \\
HD 139614   &       $<$  4.6 &        $<$  4.6 &         $<$ 3.6     &        $<$ 3.6 &         $<$ 1.2  &        $<$ 1.2  & $<$ 3.0 & $<$ 3.0 \\
HD 142527   &       $<$  9.0 &        $<$  9.0 &   6.6 $\pm$ 1.1     &  5.0 $\pm$ 2.0 &         $<$ 4.0  &        $<$ 4.0  & $<$ 4.0 & $<$ 4.0 \\
HD 144668   &       $<$  4.5 &        $<$  4.5 &         $<$ 4.8     &        $<$ 4.8 &         $<$ 0.8  &        $<$ 0.8  & $<$ 4.0 & $<$ 4.0 \\
IRS 48      &       $<$  9.6 &        $<$  9.6 &         $<$ 4.0     &        $<$ 4.0 &         $<$ 1.2  &        $<$ 1.2  & $<$ 3.0 & $<$ 3.0 \\
HD 163296   & 5.3 $\pm$  0.8 &  4.5 $\pm$  0.8 &   2.8 $\pm$ 0.5     &  2.7 $\pm$ 0.5 &  1.3  $\pm$ 0.2  & 0.74 $\pm$ 0.23 & $<$ 3.0 & $<$ 3.0 \\
HD 169142   &       $<$  8.4 &        $<$  8.4 &         $<$ 7.9     &        $<$ 7.9 &         $<$ 2.4  &        $<$ 2.4  & $<$ 4.0 & $<$ 4.0 \\
HD 179218   &       $<$  3.2 &        $<$  3.2 &         $<$ 2.7     &        $<$ 2.7 &         $<$ 1.1  &        $<$ 1.1  & $<$ 2.0 & $<$ 2.0 \\
\hline                                                                                                                      
DG Tau      & 5.9 $\pm$  0.7 &  8.9 $\pm$  0.7 &  10.6 $\pm$ 0.7     & 10.4 $\pm$ 0.7 &  2.6  $\pm$ 0.4  & 3.8  $\pm$ 0.4  & 4.7 $\pm$  0.8 &  4.7 $\pm$ 0.8  \\
AS 205      & 6.5 $\pm$  1.0 &  6.4 $\pm$  1.0 &   4.8 $\pm$ 0.8     &  4.7 $\pm$ 0.8 &  0.8  $\pm$ 0.3  & 1.3  $\pm$ 0.3  & 2.2 $\pm$  0.8 &  2.5 $\pm$ 0.8  \\  
SR 21       &       $<$  3.8 &        $<$  3.8 &         $<$ 3.9     &        $<$ 3.8 &         $<$ 1.3  &        $<$ 1.3  &       $<$  3.5 &        $<$ 3.5  \\ 
S CrA       & 5.0 $\pm$  0.6 &  5.8 $\pm$  0.6 &   4.5 $\pm$ 0.7     &  5.1 $\pm$ 0.7 & absorption       & absorption      & 2.9 $\pm$  0.8 &        $<$ 4.5  \\  
 & & & & & & \\
\hline \hline
& \multicolumn{6}{c}{$^2\Pi_{1/2}$}  & & \\
& $9/2^--7/2^+$  &  $9/2^+-7/2^-$   &  $7/2^--5/2^+$  & $7/2^+-5/2^-$   & $3/2^+-1/2^-$   & $3/2^--1/2^+$    & & \\
& 55.89\,\micron & 55.95\,\micron  &  71.17\,\micron & 71.21\,\micron & 163.12\,\micron & 163.40\,\micron & & \\
\hline
AB Aur      &       $<$ 22.4 &       $<$ 22.4 &  4.5 $\pm$ 0.6 &  4.5 $\pm$ 0.6 & $<$ 3.3 & $<$ 3.3 & & \\
HD 36112    &       $<$  7.6 &       $<$  7.6 &  1.2 $\pm$ 0.2 &  1.2 $\pm$ 0.3 & $<$ 1.5 & $<$ 1.5 & & \\
HD 38120    &       $<$  5.6 &       $<$  5.6 &        $<$ 2.3 &        $<$ 2.3 & $<$ 1.3 & $<$ 1.3 & & \\
HD 50138    &       $<$  8.0 &       $<$  8.0 &  2.0 $\pm$ 0.4 &  2.0 $\pm$ 0.4 & $<$ 1.8 & $<$ 1.8 & & \\
HD 97048    & 3.0 $\pm$  0.8 & 2.4 $\pm$  0.8 &  1.8 $\pm$ 0.4 &  1.8 $\pm$ 0.4 & $<$ 2.5 & $<$ 2.5 & & \\
HD 100453   &       $<$  5.5 &       $<$  5.5 &        $<$ 3.0 &        $<$ 3.0 & $<$ 1.4 & $<$ 1.4 & & \\
HD 100546   &       $<$ 16.0 &       $<$ 16.0 &  8.0 $\pm$ 1.4 &  8.0 $\pm$ 1.4 & $<$ 3.7 & $<$ 3.7 & & \\
HD 104237   & 2.4 $\pm$  0.8 & 2.7 $\pm$  0.8 &  1.4 $\pm$ 0.3 &  1.4 $\pm$ 0.3 & $<$ 1.7 & $<$ 1.7 & & \\
HD 135344 B &       $<$  8.2 &       $<$  8.2 &        $<$ 2.7 &        $<$ 2.7 & $<$ 1.6 & $<$ 1.6 & & \\
HD 139614   &       $<$  8.5 &       $<$  8.5 &        $<$ 2.9 &        $<$ 2.9 & $<$ 1.6 & $<$ 1.6 & & \\
HD 142527   &       $<$ 13.0 &       $<$ 13.0 &        $<$ 6.3 &        $<$ 6.3 & $<$ 2.8 & $<$ 2.8 & & \\
HD 144668   &       $<$  7.8 &       $<$  7.8 &        $<$ 2.9 &        $<$ 2.9 & $<$ 2.3 & $<$ 2.3 & & \\
IRS 48      &       $<$  8.3 &       $<$  8.3 &        $<$ 2.9 &        $<$ 2.9 & $<$ 1.6 & $<$ 1.7 & & \\
HD 163296   & 4.8 $\pm$  1.0 & 6.0 $\pm$  1.0 &        $<$ 1.8 &  2.9 $\pm$ 0.3 & $<$ 1.4 & $<$ 1.4 & & \\
HD 169142   &       $<$ 13.5 &       $<$ 13.5 &        $<$ 5.8 &        $<$ 5.8 & $<$ 2.8 & $<$ 2.8 & & \\
HD 179218   &       $<$  7.0 &       $<$  7.0 &        $<$ 2.2 &        $<$ 2.2 & $<$ 1.0 & $<$ 1.0 & & \\
\hline                                                                           
DG Tau      & 5.0 $\pm$  1.0 & 5.0 $\pm$  1.0 &  4.0 $\pm$ 1.0 &  4.0 $\pm$ 1.0 & 0.9 $\pm$ 0.4 & 1.3 $\pm$ 0.4 & & \\  
AS 205      & 3.8 $\pm$  1.2 & 5.6 $\pm$  1.2 &  2.1 $\pm$ 0.6 &  2.1 $\pm$ 0.6 & $<$ 1.5       & $<$ 1.5       & & \\
SR 21       &       $<$  5.8 &       $<$  5.8 &        $<$ 3.2 &        $<$ 3.2 & $<$ 1.5       & $<$ 1.5       & & \\
S CrA       &       $<$  6.0 &       $<$  6.0 &  1.6 $\pm$ 0.6 &  1.9 $\pm$ 0.6 & $<$ 1.6       & $<$ 1.6       & & \\
\hline\hline
\end{tabular}
\tablefoot{Units and upper limits as in Table \ref{tab:atomic}. 
\tablefoottext{a} Blended with CO $J$=31-30.} 
\end{table*}

\begin{table*}
\caption{Atomic and molecular data of the far-IR detected transitions}
\label{tab:moldata}
\centering
\begin{tabular}{llrrr}
\hline\hline
Species      & Transition & $\lambda$  & $E_{u}$ & log($A_{ul}$)  \\ 
             &            & [\micron]     & [K]    & [s$^{-1}$]     \\ 
\hline
OH              & $^2\Pi_{1/2} \, 9/2^+ - 7/2^-$ & 55.891  & 875 & 0.34 \\
OH              & $^2\Pi_{1/2} \, 9/2^- - 7/2^+$ & 55.949  & 875 & 0.34 \\
OH              & $^2\Pi_{3/2} \, 9/2^- - 7/2^+$ & 65.131  & 512 & 0.11 \\
OH              & $^2\Pi_{3/2} \, 9/2^+ - 7/2^-$ & 65.278  & 510 & 0.10 \\
OH              & $^2\Pi_{1/2} \, 7/2^- - 5/2^+$ & 71.170  & 617 & 0.01 \\
OH              & $^2\Pi_{1/2} \, 7/2^+ - 5/2^-$ & 71.215  & 617 & 0.01 \\
OH              & $^2\Pi_{1/2} \ - \ ^2\Pi_{3/2} \ J = 1/2^- - 3/2^+$ & 79.115 & 181 & -1.44 \\
OH              & $^2\Pi_{1/2} \ - \ ^2\Pi_{3/2} \ J = 1/2^+ - 3/2^-$ & 79.178 & 181 & -1.44 \\
OH              & $^2\Pi_{3/2} \, 7/2^- - 5/2^+$ & 84.420  & 291 & -0.28  \\
OH              & $^2\Pi_{3/2} \, 7/2^+ - 5/2^-$ & 84.596  & 290 & -0.28  \\
OH              & $^2\Pi_{3/2} \, 5/2^- - 3/2^+$ & 119.233 & 120 & -0.86  \\
OH              & $^2\Pi_{3/2} \, 5/2^+ - 3/2^-$ & 119.441 & 120 & -0.86  \\
OH              & $^2\Pi_{1/2} \, 3/2^+ - 1/2^-$ & 163.120 & 270 & -1.190 \\
OH              & $^2\Pi_{1/2} \, 3/2^- - 1/2^+$ & 163.410 & 270 & -1.190 \\
\hline
CH$^+$          & $J = 6-5$ & 60.248  & 839  & 0.27  \\
CH$^+$          & $J = 5-4$ & 72.141  & 600  & 0.03  \\
CH$^+$          & $J = 4-3$ & 90.017  & 400  & -0.26 \\
CH$^+$          & $J = 3-2$ & 119.858 & 240  & -0.66 \\
CH$^+$          & $J = 2-1$ & 179.605 & 120  & -1.21 \\
\hline
o-H$_2$O        & $4_{32}-3_{21}$ &  58.70 &    550 &  0.14 \\
p-H$_2$O        & $7_{26}-6_{15}$ &  59.99 &   1020 &  0.13 \\
o-H$_2$O        & $8_{18}-7_{07}$ &  63.32 &   1070 &  0.24 \\
o-H$_2$O        & $7_{16}-6_{25}$ &  66.09 &   1013 & -0.02 \\
o-H$_2$O        & $3_{30}-2_{21}$ &  66.44 &    410 &  0.09 \\             
o-H$_2$O        & $7_{07}-6_{16}$ &  71.95 &    843 &  0.06 \\
o-H$_2$O        & $3_{21}-2_{12}$ &  75.38 &    305 & -0.48 \\
o-H$_2$O        & $4_{23}-3_{12}$ &  78.74 &    432 & -0.32 \\
p-H$_2$O        & $6_{15}-5_{24}$ &  78.93 &    781 & -0.34 \\
o-H$_2$O        & $6_{16}-5_{05}$ &  82.03 &    643 &  0.06 \\
p-H$_2$O        & $6_{06}-5_{15}$ &  83.28 &    642 & -0.15 \\
o-H$_2$O        & $2_{21}-1_{10}$ & 108.07 &    194 & -0.59 \\
o-H$_2$O        & $4_{14}-3_{03}$ & 113.54 &    323 & -0.61 \\
o-H$_2$O        & $2_{12}-1_{01}$ & 179.53 &    114 & -1.25 \\
p-H$_2$O        & $4_{13}-4_{04}$ & 187.11 &    396 & -1.43 \\
\hline
$[\ion{O}{i}]$  & $^3P_1 \ - \ ^3P_2$               & 63.185  & 228 & -4.05 \\
$[\ion{O}{i}]$  & $^3P_0 \ - \ ^3P_2$               & 145.535 & 327 & -4.75   \\
\hline
$[\ion{C}{ii}]$ & $^2P_{3/2} \ - \ ^2P_{1/2}$         & 157.680 & 91  & -5.64  \\
\hline\hline
\end{tabular}
\end{table*}

\section{Effects of non-LTE excitation of OH far-IR}\label{ap:nlte}
\begin{figure*}
\centering
\includegraphics[width=0.4\hsize]{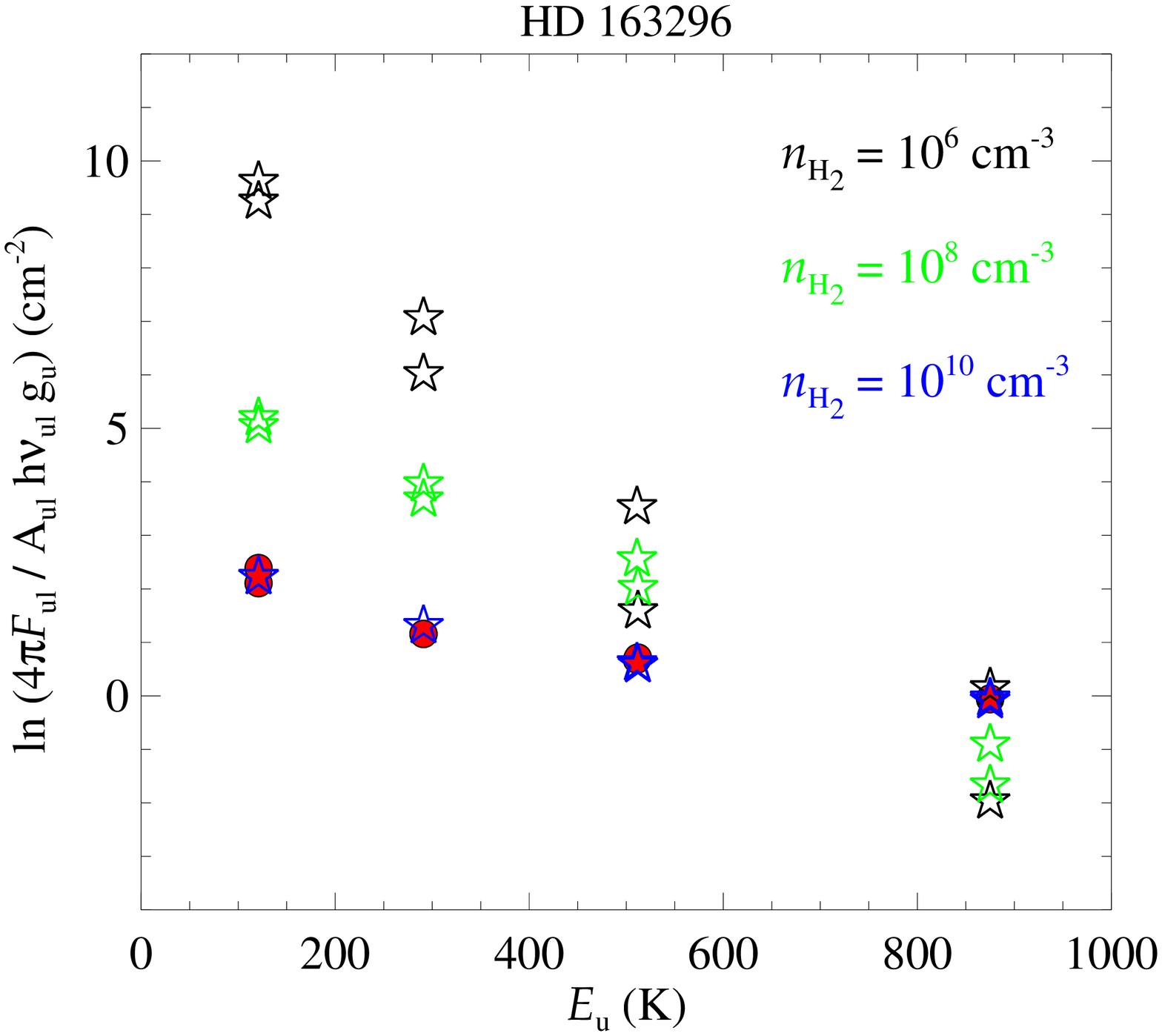}
\includegraphics[width=0.4\hsize]{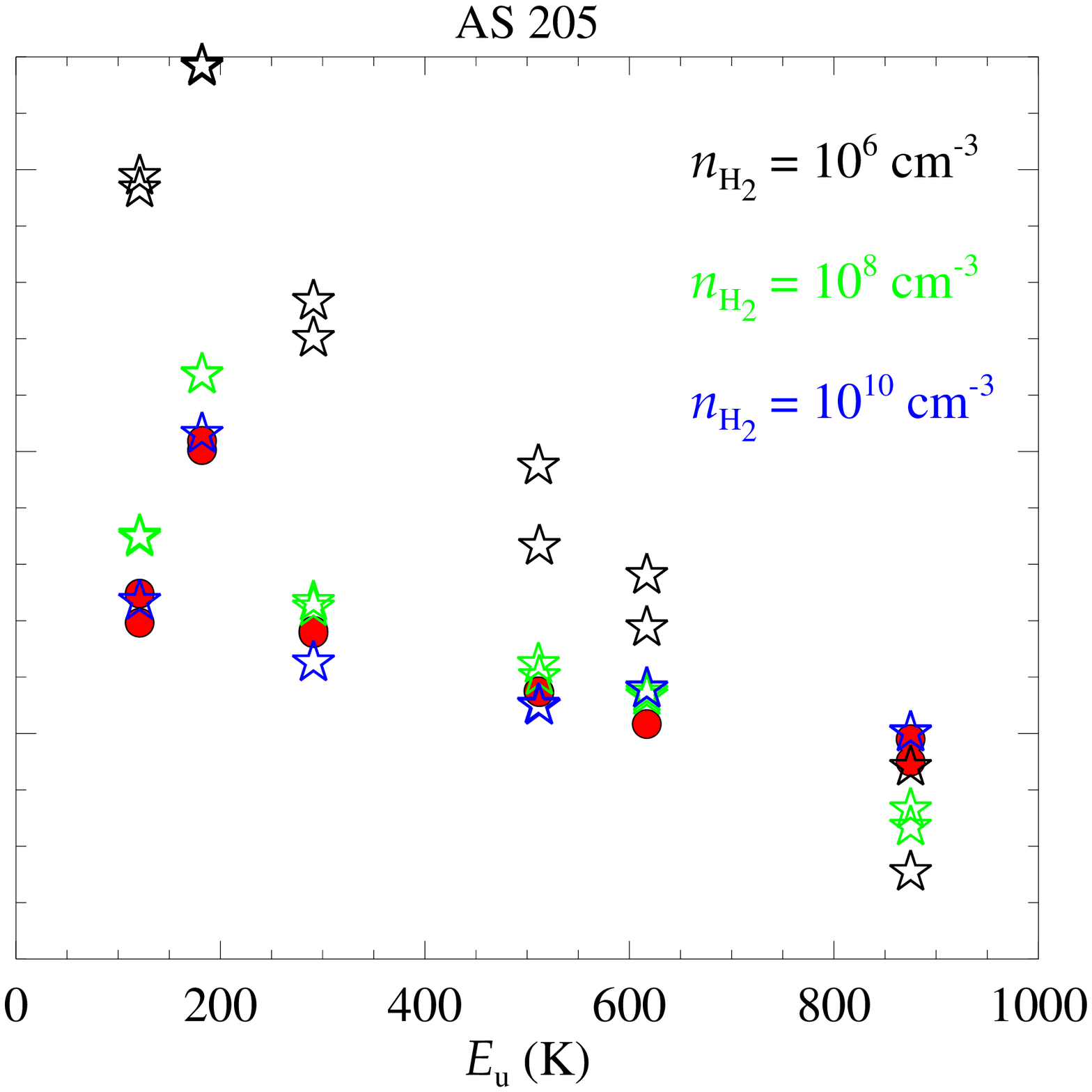}
\includegraphics[width=0.4\hsize]{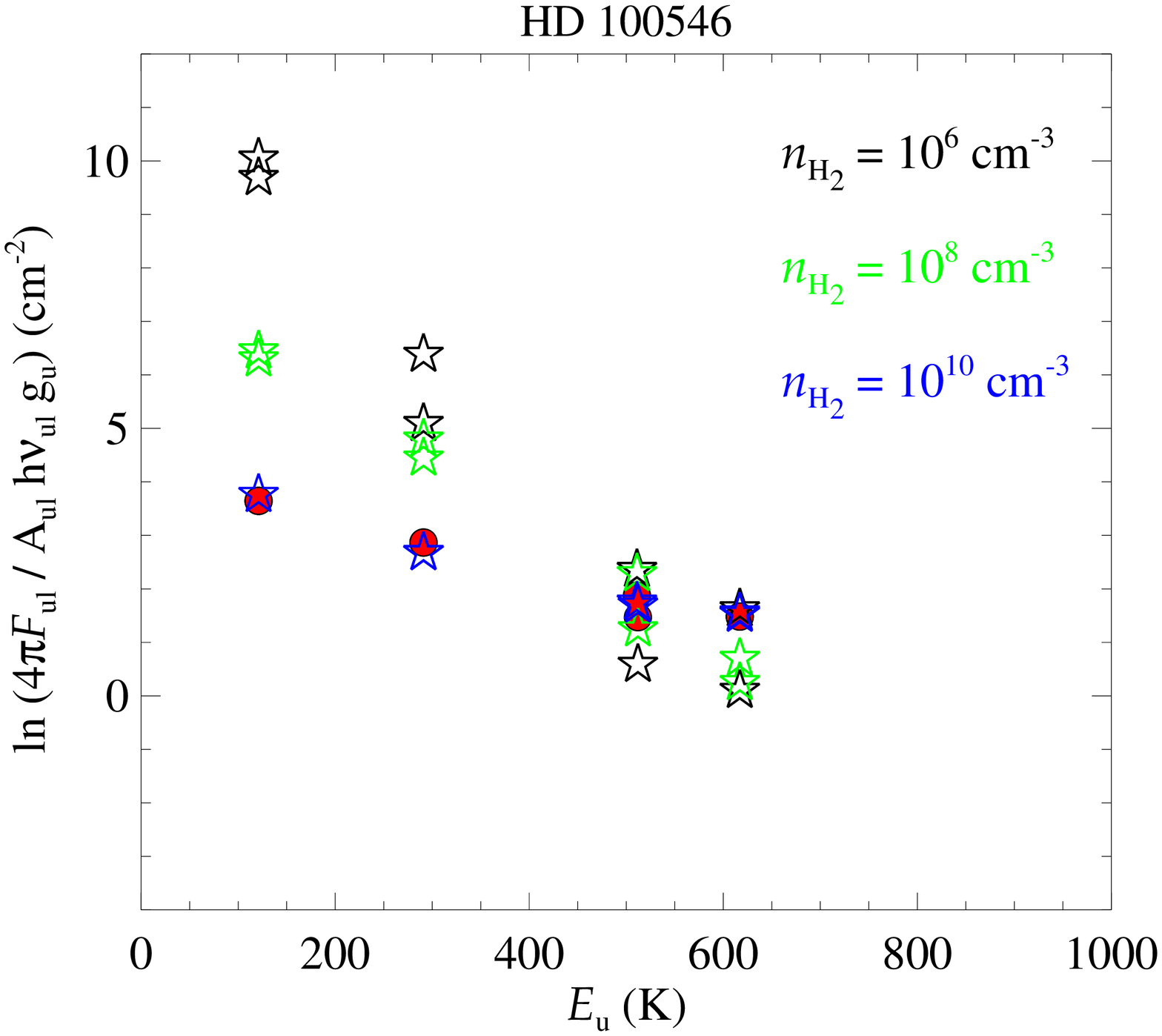}
\includegraphics[width=0.4\hsize]{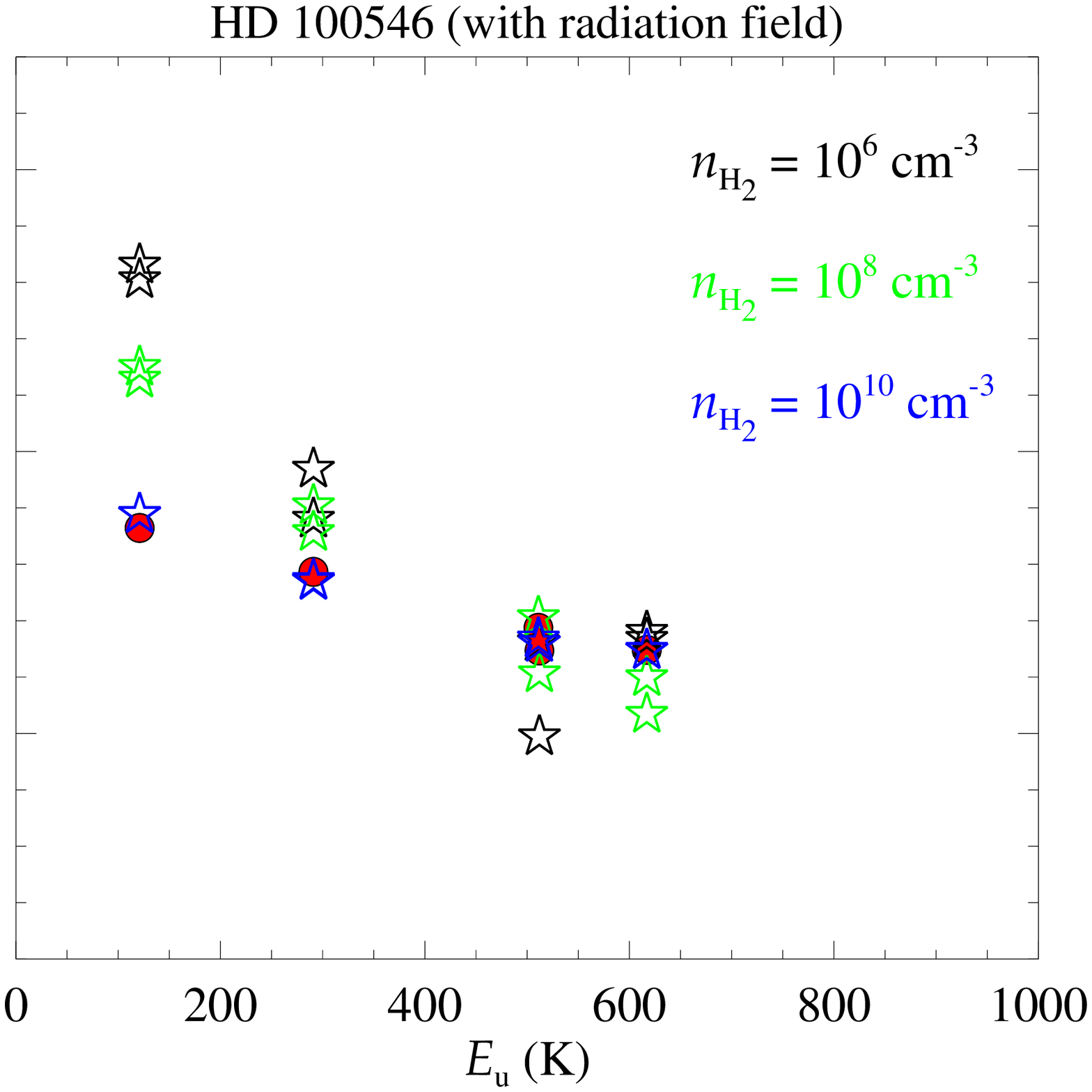}
\includegraphics[width=0.4\hsize]{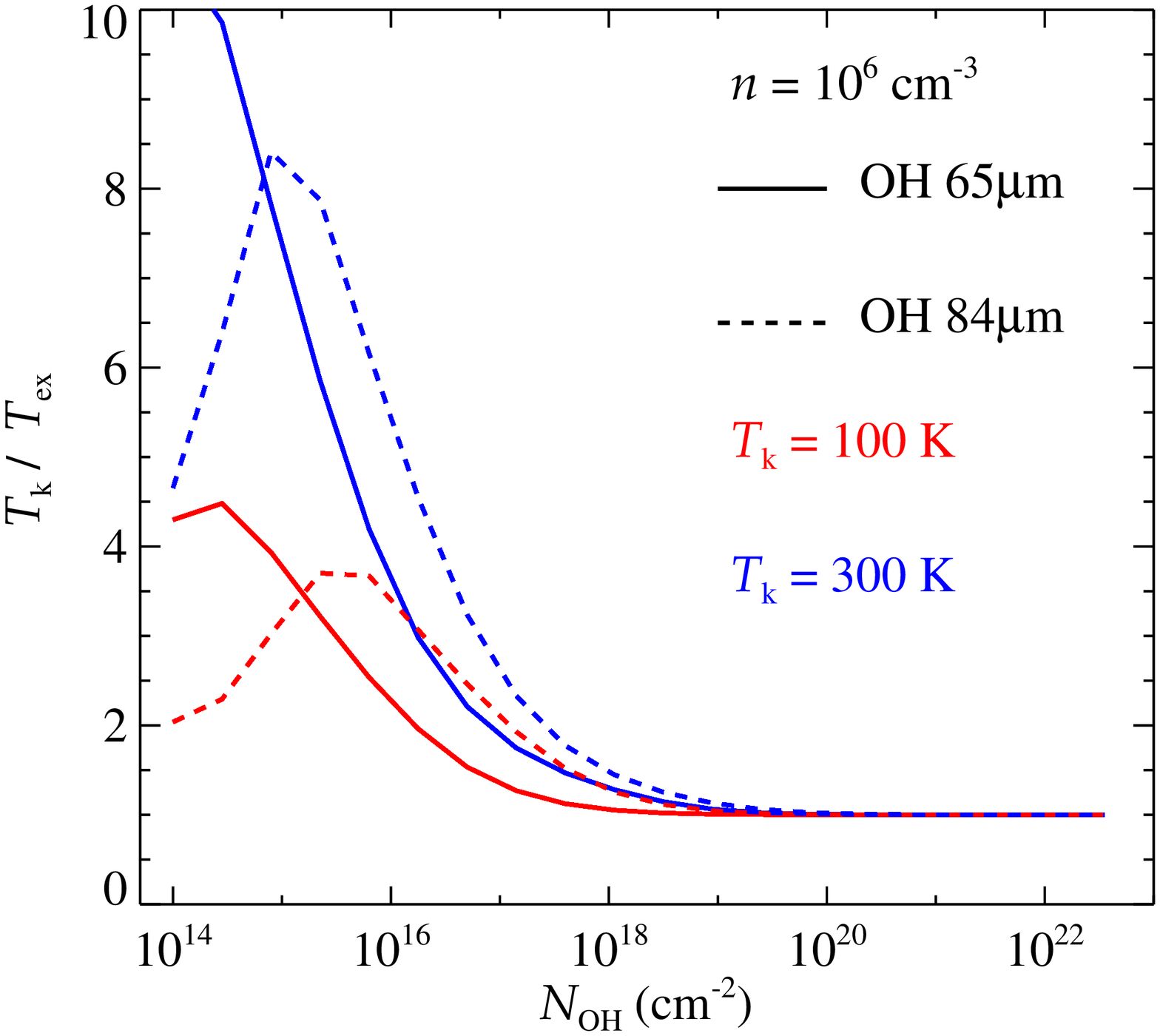}
\includegraphics[width=0.4\hsize]{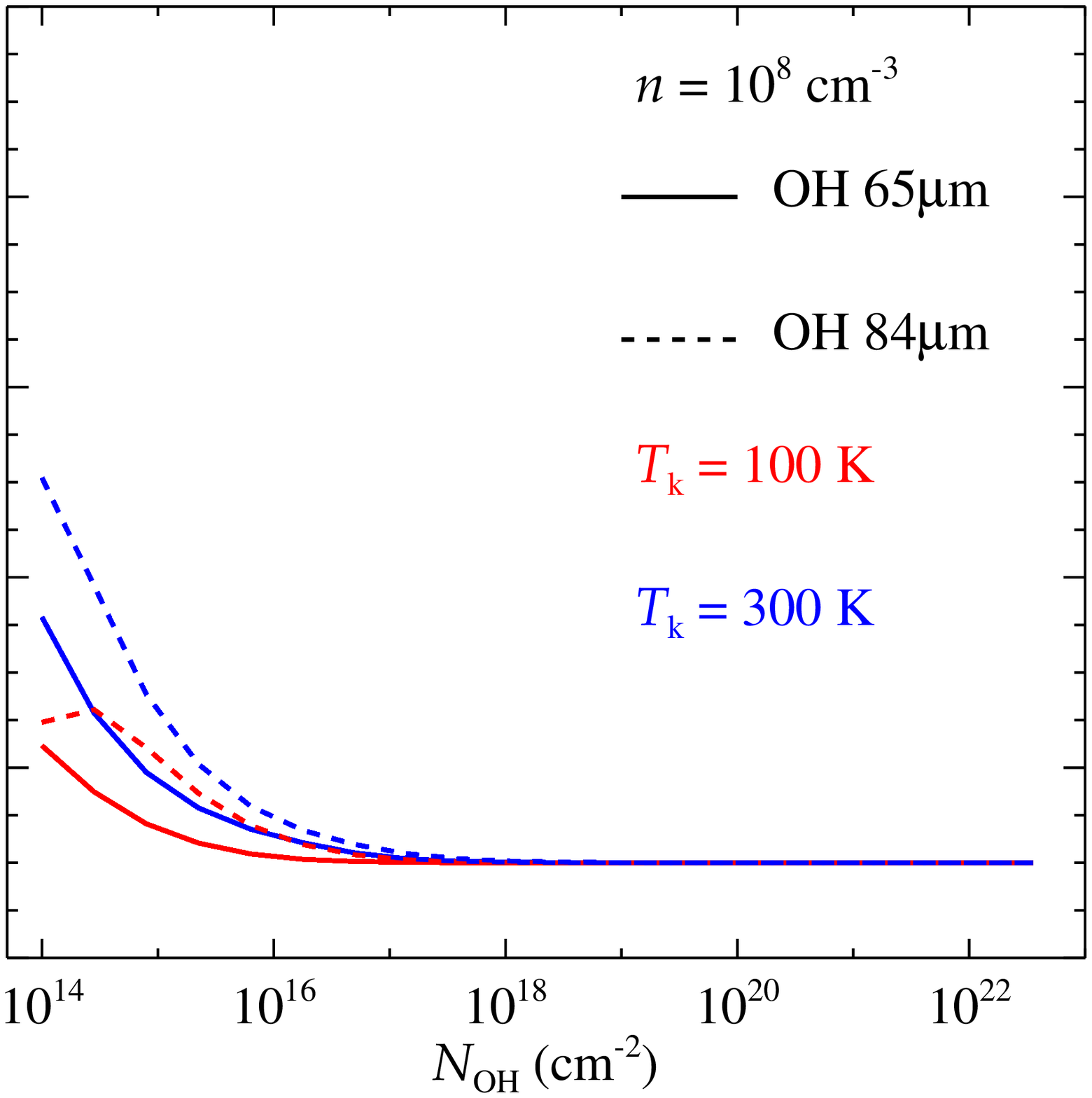}
\caption{Results of the non-LTE simulations with RADEX. ({\it top}) OH rotational diagram of HD 163296 and 
AS 205 and non-LTE model predictions: three different models are shown for different values of the gas density and temperature
of 400\,K and 200\,K for HD 163296 and AS 205, respectively. 
Only models with $n \geq 10^{10}\,$cm$^{-3}$ can reproduce the observations. 
({\it middle}) OH rotational diagram of HD 100546 and non-LTE model predictions ($N = 2 \times 10^{14}$\,\column , $T = 200\,$K) in 
the case in which the infrared radiation field is included in the RADEX simulation to test the effect of infrared pumping: in both 
cases (with and without radiation field) high gas densities are needed to reproduce the observed rotational diagram.
({\it bottom}) Ratio of $T_{\rm k}$ to $T_{\rm ex}$ for 2 OH transitions as a function of $N_{\rm OH}$ in two low gas density cases. 
Even in the low gas density cases, the OH rotational levels are in LTE ($T_{\rm k} = T_{\rm ex}$) for large values of $N_{\rm OH}$}
when the lines are optically thick.
\label{fig:nlte}
\end{figure*}

 OH lines studied here have large critical densities ($n_{\rm crit} \sim 10^9 - 10^{10}\,$cm$^{-3}$) and non-LTE
excitation may be important if the gas density is not high enough to thermalize the OH molecules. 
To test the assumption of LTE we fit the observed rotational diagram using the non-LTE 
code RADEX \citep{vdTak07}. We used the same fitting procedure as for the LTE case and we repeated 
the analysis for different values of the gas (H$_2$, collision partner) density ($n =10^6, 10^8, 
10^{10}\,$cm$^{-3}$). Fig.~\ref{fig:nlte} shows the OH rotational diagram for two test cases: the Herbig 
Ae HD 163296 and the T Tauri AS 205 disks. We reproduced the slab model using the best-fit parameters found 
in the LTE case with $N_{\rm OH} = 10^{15}\,$\column \ and $T_{\rm K} = 400\,$K  for HD 163296 and 
$N_{\rm OH} = 8 \times 10^{15}\,$\column \ and $T_{\rm K} = 200\,$K for AS 205 (Table~\ref{tab:slab}). 
The non-LTE model predictions are plotted 
in Fig.~\ref{fig:nlte} with different colors for the three values of $n_{\rm H_2}$. For low $n_{\rm H_2}$ values 
($\leq 10^8$\,cm$^{-3}$) the model fails to reproduce the observed rotational diagram. The gas density must be 
$n_{\rm H_2} \geq 10^{10}$\,cm$^{-3}$ in order to fit the observations. Thus, the OH rotational lines emerge from an 
high density region where the OH molecules are thermalized and the rotational levels are in LTE. 

\smallskip
\noindent
Infrared pumping can be relevant for the excitation of OH molecules. To test the effects of line pumping 
we run a grid of RADEX models for HD 100546 providing also the infrared radiation field 
(between 20\,\micron \ -- 3\,mm) in the input parameters. The radiation field is taken from the full disk 
thermo-chemical model of \citet{Bruderer12} 
who computed the radiation field at each position of the disk for different wavelengths. The radiation field is stronger
in the inner region of the disk ($r < 20\,$AU). As input to RADEX we considered the value of the infrared radiation 
field at a distance of $r$=20\,AU and height above the midplane $z$=4\,AU ($z/r$ = 0.2). At larger radii and height 
(where the FIR OH lines originate) the radiation field is always fainter. Fig.~\ref{fig:nlte} shows the OH rotational 
diagram of HD 100546 (middle row) and the RADEX predictions without (left) and with (right) infrared radiation field. 
The line flux ratios vary in the presence of infrared pumping, but even in this case high gas density 
($\geq 10^{10}\,$cm$^{-3}$) is needed to reproduce the observed rotational diagram.   

\begin{figure}[!ht]
\centering
\includegraphics[width=0.9\hsize]{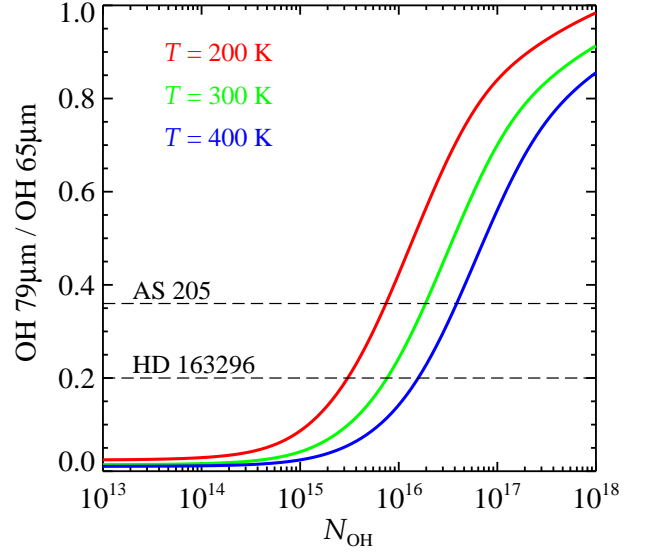}
\caption{Ratio of the OH 79\,\micron \ to the OH 65\,\micron \ lines from the LTE calculation at 3 different temperatures. The ratio increases rapidly with column density. The dashed lines indicate the observed ratio for AS 205 and HD 163296 (1\,$\sigma$ upper limit).}
\label{fig:oh-lineratio}
\end{figure}

\smallskip
\noindent
The non-LTE simulations also show that for large values of the column density ($N_{\rm OH} \geq 10^{18}\,$\column) 
the OH rotational lines are in LTE even at gas densities $\leq 10^8\,$cm$^{-3}$ (Fig.~\ref{fig:nlte}, bottom). This 
is due to line opacity which traps the photons and helps to thermalize the gas. However, we can exclude this scenario 
for most of the sources based on the non-detection of the intra-ladder transitions at 79\,\micron. These transitions 
are indeed very sensitive 
to line opacity and the lines are easily detected for $N_{\rm OH} \gtrsim 10^{16}\,$\column, as in the case of DG Tau
and AS 205. This is shown in Fig.~\ref{fig:oh-lineratio} where the ratio of the OH 79\,\micron \ to the OH 65\,\micron
\ lines is shown in the LTE case for different temperatures. In order for the intra-ladder lines to be detected high
column density is needed.
  
\section{{\rm [\ion{O}{i}], [\ion{C}{ii}]} spatial extent}{\label{ap:cii}}
This section describes an analysis of the atomic lines aiming at addressing
the spatial extent of the line. The Herschel/PACS PSF varies substantially from 50\,\micron \
to 200\,\micron. As a consequence the amount of flux in the central spaxel 
varies from $\sim$ 70\,\% at 60\,\micron \ to 55\,\% at 160\,\micron. 
For this reason, line emission can be detected outside the central spaxel (especially in the red part $>$ 100\,\micron).
To check whether a line is spatially extended we compute the equivalent width ($W$) and 
integrated continuum ($F_{c}$) next to the line and check the relative spatial distribution. 
If the line emission is co-spatial to the continuum emission, then the spatial distribution of the equivalent
width  will be equal to that of the integrated continuum ($F_c$) (same PSF). In particular, the
distribution of $F_c$ corresponds to the PSF at the given wavelength (assuming that the continuum emission
is not spatially resolved). 

\begin{figure*}
\centering
\includegraphics[bb = 28 0 622 600,width=0.4\hsize]{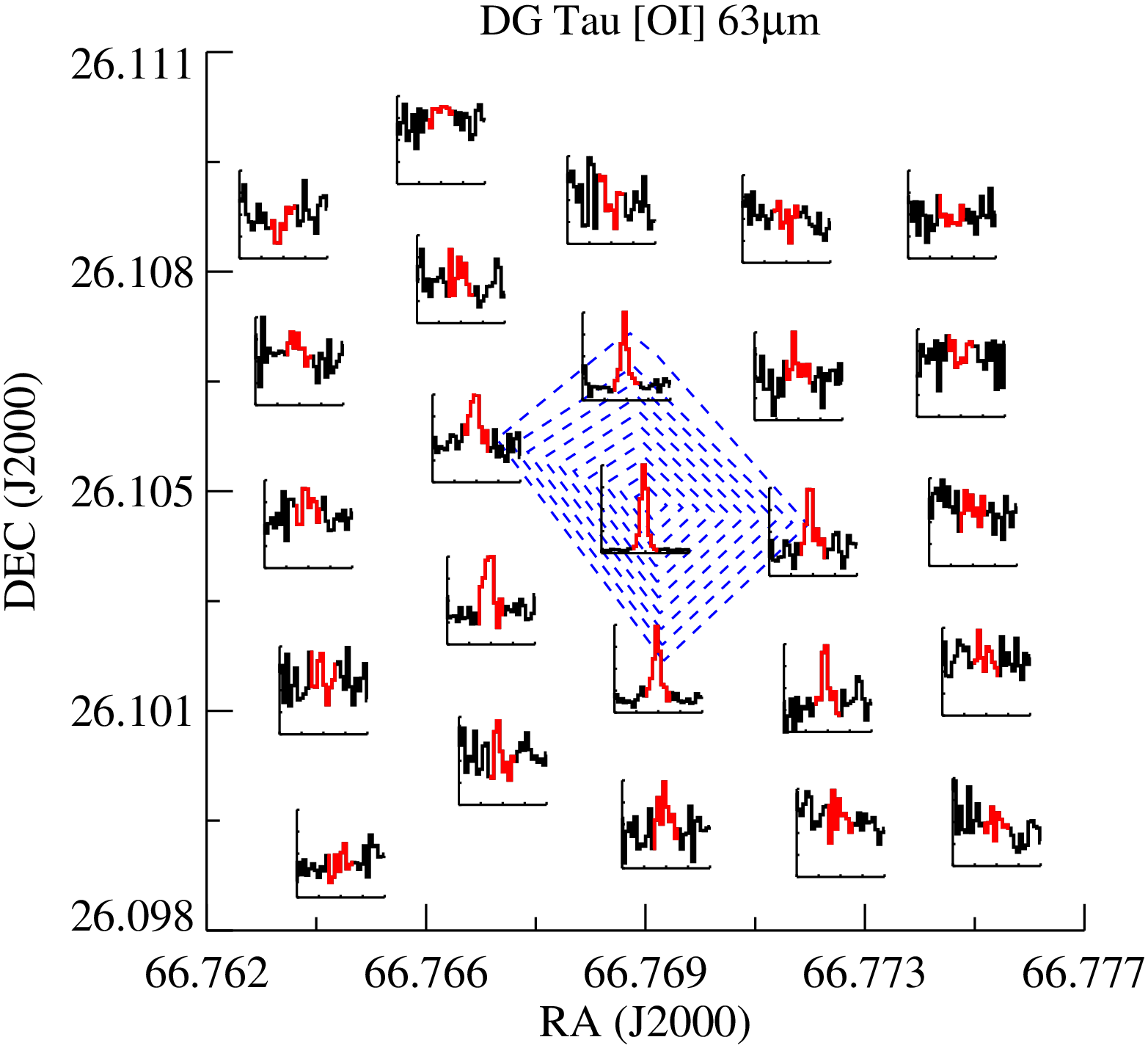}
\includegraphics[bb =  0 0 594 600,width=0.4\hsize]{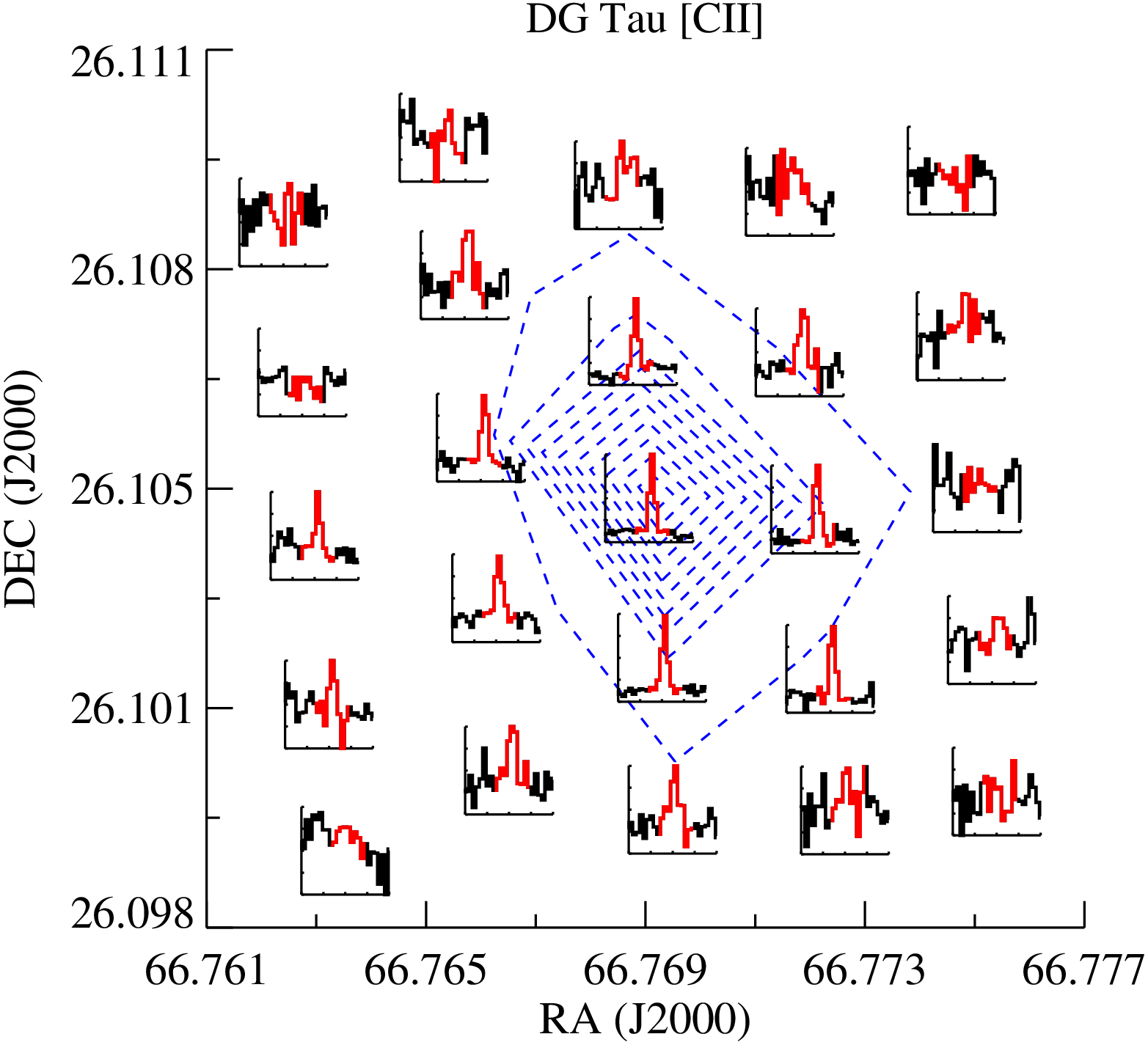}
\caption{\oia \ and \cii \ spectral map in DG Tau showing the spatially extended line emission. The contours represent 
the spectral continuum measured in the vicinity of the line, the last contour level corresponds to 10\% of the 
continuum peak. The sub-panels show the line spectrum measured in each spaxel.}\label{fig:map1}
\end{figure*}

\smallskip
\noindent
For the \oia \ line, $F_C$ is measured integrating the spectrum between 64.0 -- 64.5\,\micron \ and the equivalent 
width is measured integrating the spectrum between 63.08 -- 63.30\,\micron. The only source where off-source oxygen 
excess emission is DG Tau. Fig.~\ref{fig:map1} shows the \oia \ spectral map of DG Tau: the (blue) dashed contours 
show the distribution of the spectral continuum and the sub-panels shows the \oia \ spectrum in each spaxel. While the
continuum is compact and centered on the central spaxel, the line emission shows an excess emission outside the 
central spaxel. The maximum excess is measured southward of the central source in agreement with the outflow position.

\begin{figure*}
\centering
\includegraphics[bb = 28 0 622 600,width=0.4\hsize]{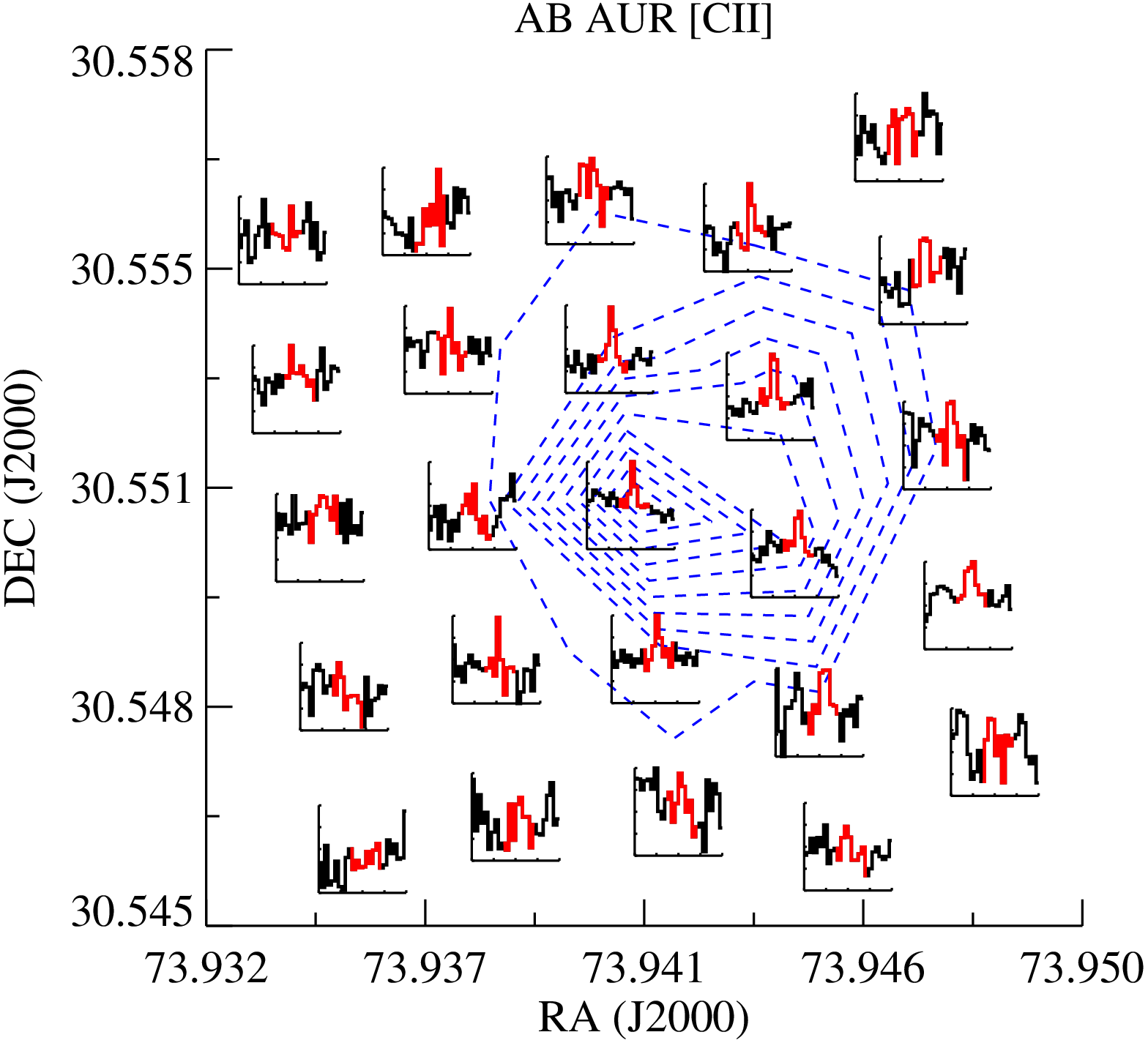}
\includegraphics[bb =  0 0 594 600,width=0.4\hsize]{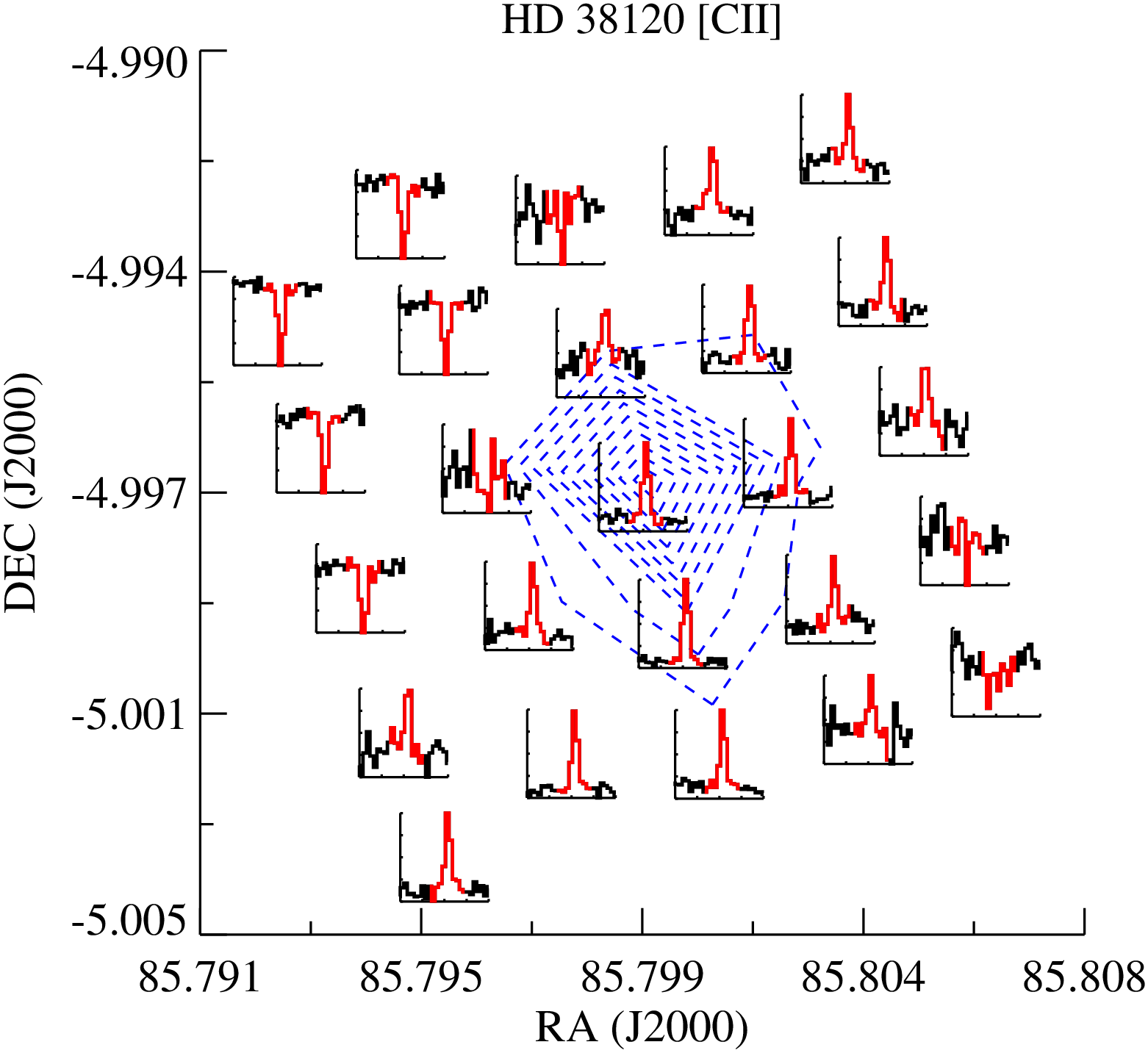}
\includegraphics[bb = 28 0 622 600,width=0.4\hsize]{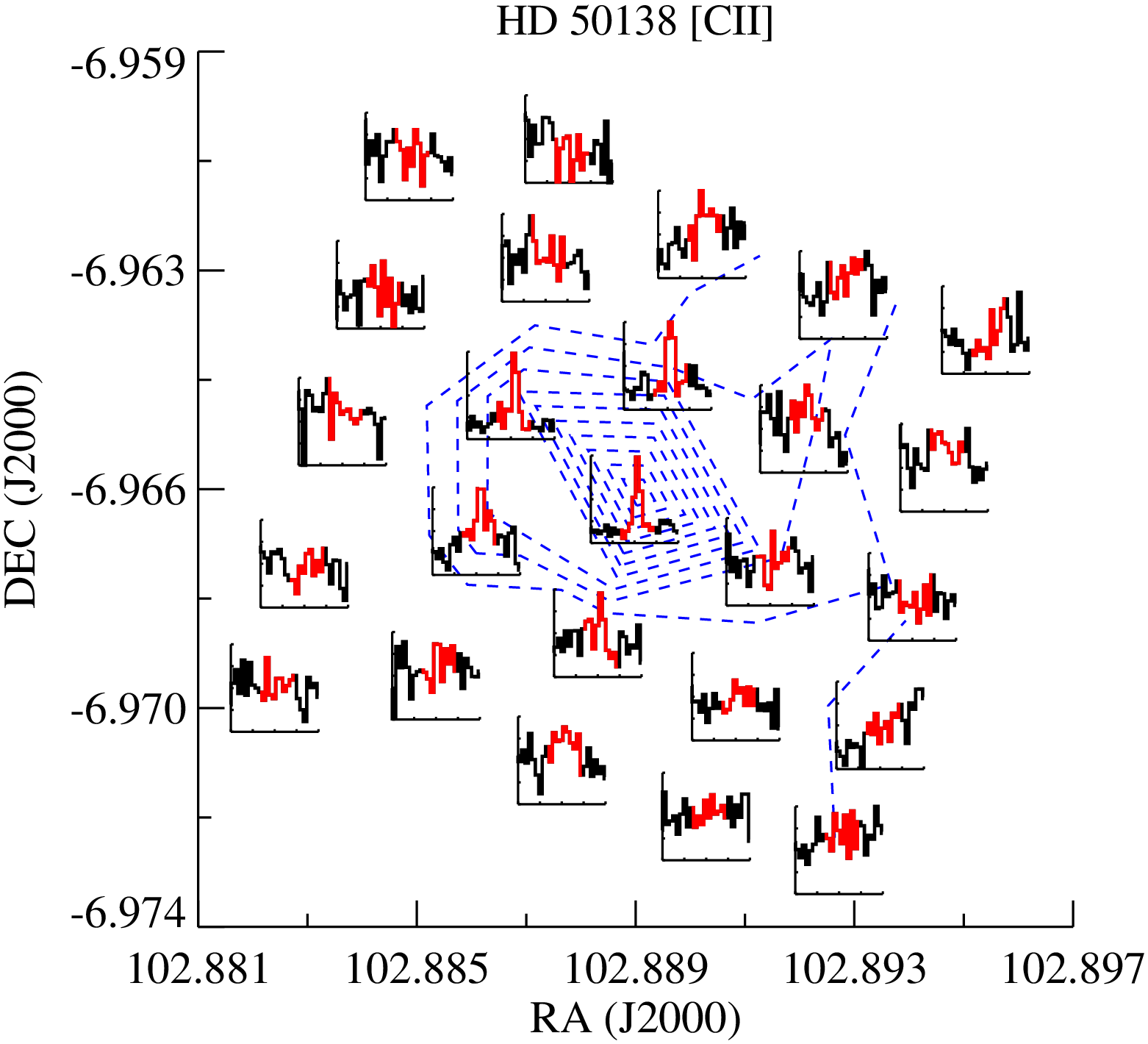}
\includegraphics[bb =  0 0 594 600,width=0.4\hsize]{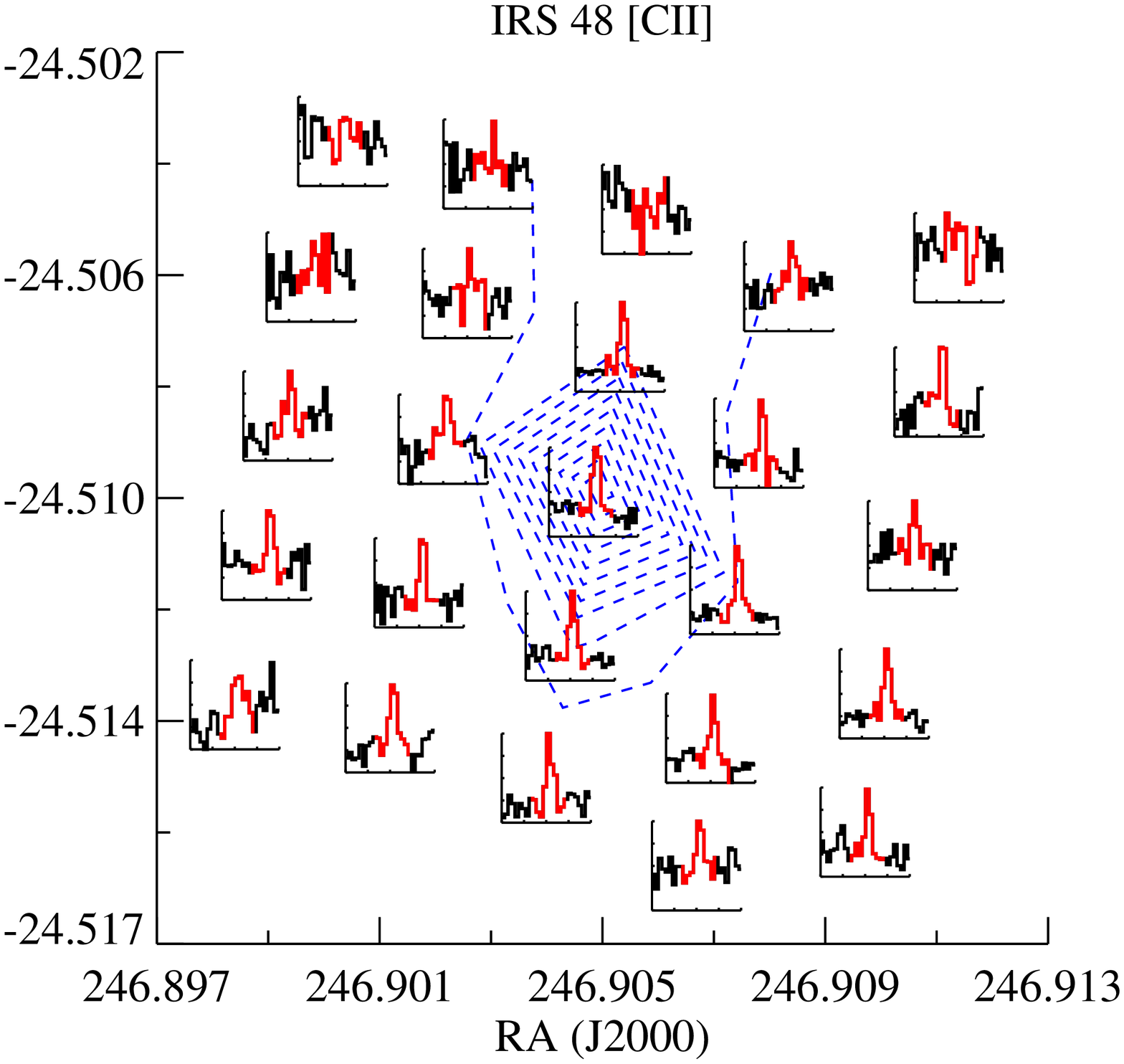}
\includegraphics[bb = 28 0 622 600,width=0.4\hsize]{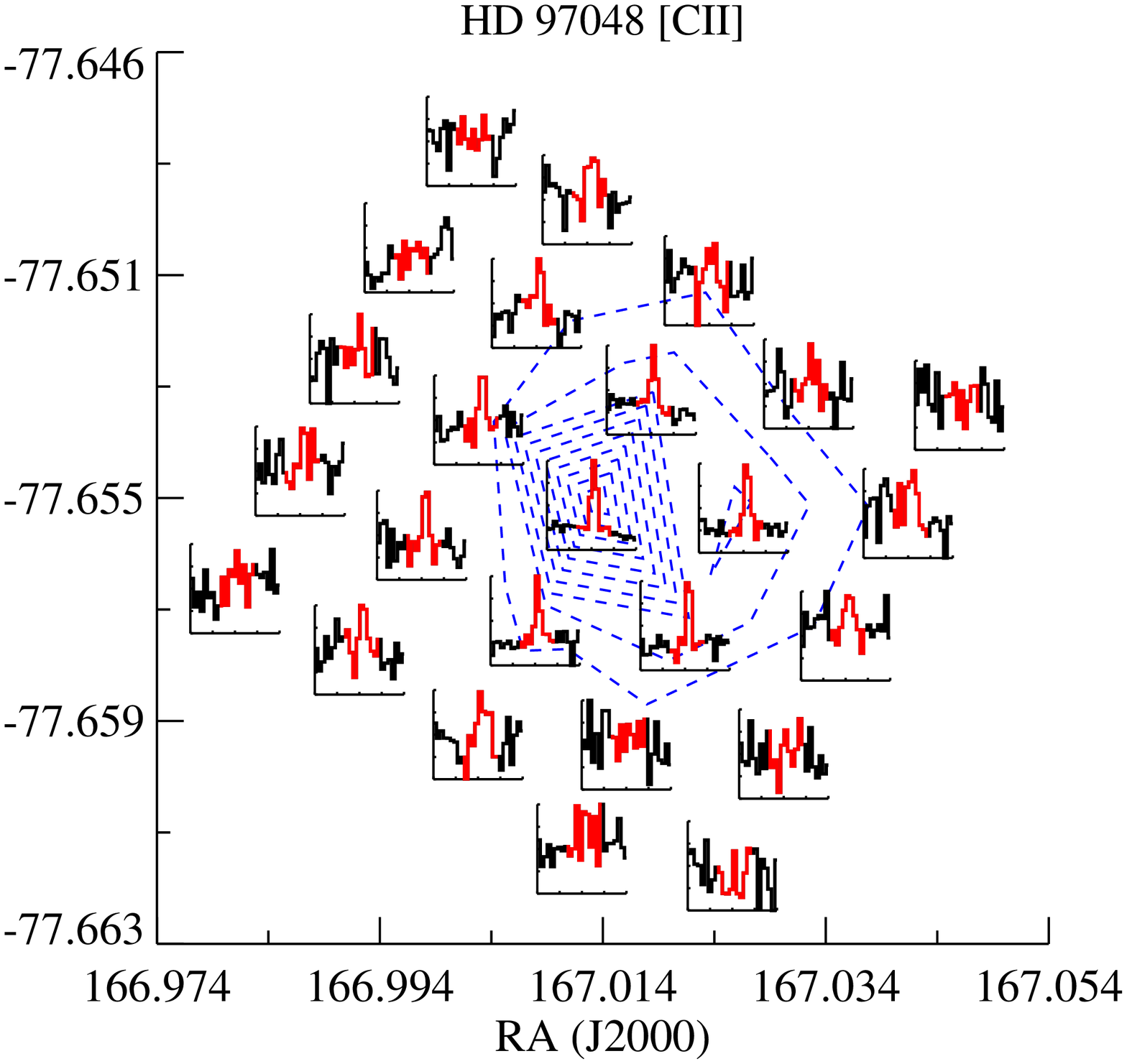}
\includegraphics[bb =  0 0 594 600,width=0.4\hsize]{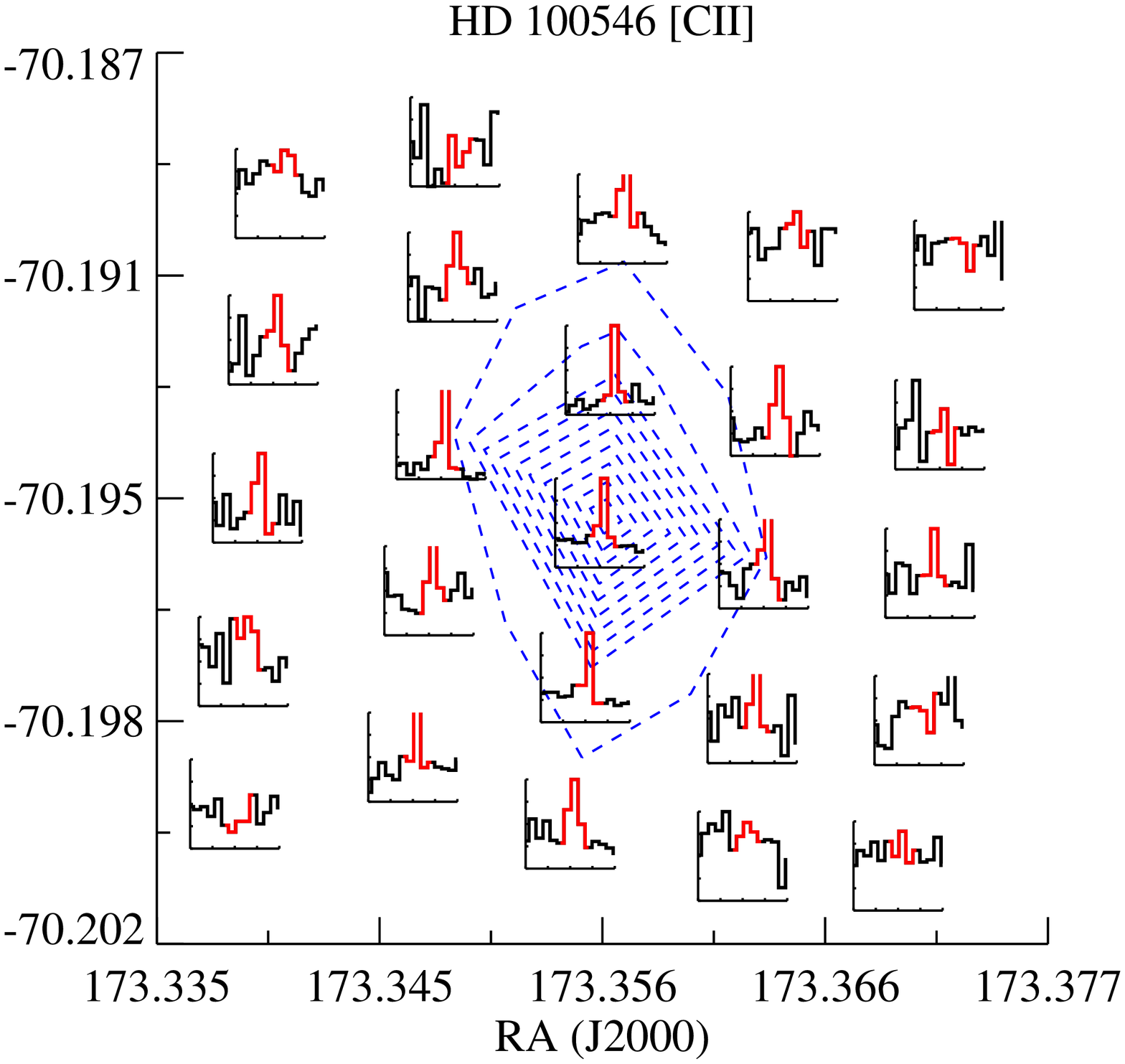}
\caption{Same as Fig.~\ref{fig:map1} for \cii \ emission in Herbig AeBe sources.}\label{fig:map2}
\end{figure*}

\smallskip
\noindent
For the \cii \ line the line is integrated between 157.530 -- 157.970\,\micron \ and the continuum flux between 
158.5 -- 162\,\micron. Figs.~\ref{fig:map1} and \ref{fig:map2} shows the line spectral map for different sources. 
The spectral map shows the spectrum (continuum subtracted) in each spaxel. 
All the sources where \cii \ emission is detected show excess line emission outside the central spaxel. The most clear 
cases are HD 38120, IRS 48 and DG Tau. This pattern is the result of extended line emission. In the case of AB Aur and 
HD 97048 the object is mis-pointed and the spatial distribution of the continuum emission deviates from the PSF. 
Nevertheless, also in these two cases the line emission is not co-spatial with the continuum emission and proves a 
spatially extended line emission. The case of HD 50138 is less clear. 

\subsection{On-source {\rm \cii}  line flux}
To estimate the maximum \cii \ emission associated with the protoplanetary disk the extended emission needs to be 
subtracted.  To do this the \cii \ line flux (integrated between 157.60-157.98\,\micron) in each of the 9 central 
spaxels is calculated. Then the extended emission is determined as the average of the line flux measured in the 8 
neighbouring spaxels (around the central one) and subtracted from the value measured in the central spaxel. The 
result is reported in Table \ref{tab:atomic}. In this way, the large scale ($>$ 9\farcs4) \cii \ emission is 
approximately removed. The value of the \cii \ flux derived by this method must be considered an upper limit to the 
\cii \ emission arising from the disk as extended emission from a compact remnant envelope may still be present in 
the central 9\farcs4 $\times$ 9\farcs4 area of the sky.

\end{appendix}
\end{document}